\documentclass[10pt,journal,compsoc]{IEEEtran}
\usepackage{amsmath,amssymb}
\usepackage{colortbl}
\usepackage{diagbox}
\usepackage{subeqnarray}
\usepackage{cases}
\usepackage{setspace}
\usepackage{overpic}
\usepackage{rotating}
\usepackage{tabularx}
\usepackage{booktabs}
\usepackage{multicol}
\usepackage{dcolumn}
\usepackage{graphicx}
\usepackage{subfigure}
\usepackage{float}
\usepackage{epsfig}
\usepackage{amsmath}
\usepackage{amssymb}
\usepackage{array}
\usepackage{epstopdf}
\usepackage{color}
\usepackage{xcolor}
\usepackage[nocompress]{cite}
\usepackage{psfrag}
\usepackage{cite}
\usepackage{multirow}
\usepackage{color}
\usepackage[linesnumbered,ruled]{algorithm2e}

\graphicspath{{figures/}}

\usepackage{array}
\newcolumntype{I}{!{\vrule width 3pt}}
\newlength\savedwidth

\newlength\savewidth
\newcommand\shline{\noalign{\global\savewidth\arrayrulewidth
                            \global\arrayrulewidth 0.8pt}%
                   \hline
                   \noalign{\global\arrayrulewidth\savewidth}}

\ifCLASSOPTIONcompsoc
  \usepackage[nocompress]{cite}
\else
  \usepackage{cite}
\fi
\ifCLASSINFOpdf
\else
\fi

\usepackage[bookmarks,colorlinks,linkcolor=red, anchorcolor=blue, citecolor=green]{hyperref}
\begin{document}

\title{Plug-and-Play Image Restoration with\\ Deep Denoiser Prior}

%I.E.EEcompsocitemizethanks is a special \thanks that produces the bulleted
% affiliations. Use I.E.EEcompsocthanksitem which works much like \item
% I.E.EEcompsocitemizethanks becomes like \thanks and
% I.E.EEcompsocthanksitem becomes a line break with idention. This

\author{Kai~Zhang,
        Yawei~Li,
        Wangmeng Zuo,~\IEEEmembership{Senior Member,~IEEE,}
        Lei Zhang,~\IEEEmembership{Fellow,~IEEE,}\\
        Luc Van Gool and~Radu Timofte,~\IEEEmembership{Member,~IEEE}

\thanks{K. Zhang, Y. Li and R. Timofte are with the Computer Vision Lab, ETH Z\"{u}rich, Z\"{u}rich, Switzerland (e-mail:
kai.zhang@vision.ee.ethz.ch;
yawei.li@vision.ee.ethz.ch;
radu.timofte@vision.ee.ethz.ch).}
\thanks{L. Van Gool is with the Computer Vision Lab, ETH Z\"{u}rich, Z\"{u}rich, Switzerland, and also with KU Leuven, Leuven, Belgium (e-mail:
vangool@vision.ee.ethz.ch).}
\thanks{W. Zuo is with the School of Computer Science and
Technology, Harbin Institute of Technology, Harbin, China (e-mail:
cswmzuo@gmail.com).}
\thanks{L. Zhang is with the Department of Computing, The Hong Kong
Polytechnic University, Hong Kong, China (e-mail: cslzhang@comp.polyu.edu.hk).}
}

\IEEEtitleabstractindextext{
\begin{abstract}
Recent works on plug-and-play image restoration have shown that a denoiser can implicitly serve as the image prior for model-based methods to solve many inverse problems. Such a property induces considerable advantages for plug-and-play image restoration (e.g., integrating the flexibility of model-based method and effectiveness of learning-based methods) when the denoiser is discriminatively learned via deep convolutional neural network (CNN) with large modeling capacity. However, while deeper and larger CNN models are rapidly gaining popularity, existing plug-and-play image restoration hinders its performance due to the lack of suitable denoiser prior.
In order to push the limits of plug-and-play image restoration, we set up a benchmark deep denoiser prior by training a highly flexible and effective CNN denoiser. We then plug the deep denoiser prior as a modular part into a half quadratic splitting based iterative algorithm to solve various image restoration problems. We, meanwhile, provide a thorough analysis of parameter setting, intermediate results and empirical convergence to better understand the working mechanism. Experimental results on three representative image restoration tasks, including deblurring, super-resolution and demosaicing, demonstrate that the proposed plug-and-play image restoration with deep denoiser prior not only significantly outperforms other state-of-the-art model-based methods but also achieves competitive or even superior performance against state-of-the-art learning-based methods. The source code is available at \url{https://github.com/cszn/DPIR}.
\end{abstract}

\begin{IEEEkeywords}
Denoiser Prior, Image Restoration, Convolutional Neural Network, Half Quadratic Splitting, Plug-and-Play
\end{IEEEkeywords}}
\maketitle
\IEEEdisplaynontitleabstractindextext
% I.E.EEdisplaynontitleabstractindextext has no effect when using

% \fi
\IEEEpeerreviewmaketitle

\IEEEraisesectionheading{\section{Introduction}\label{sec:introduction}}

\IEEEPARstart{I}{mage restoration} (IR) has been a long-standing problem for its highly practical value in various low-level vision applications~\cite{richardson1972bayesian,andrews1977digital}.
In general, the purpose of image restoration is to recover the latent clean image $\mathbf{x}$ from its degraded observation $\mathbf{y} = \mathcal{T}(\mathbf{x}) + \mathbf{n}$, where $\mathcal{T}$ is the noise-irrelevant degradation operation, $\mathbf{n}$ is assumed to be additive white Gaussian noise (AWGN) of standard deviation $\sigma$.
By specifying different degradation operations, one can correspondingly get different IR tasks. Typical IR tasks would be image denoising when $\mathcal{T}$ is an identity operation, image deblurring when $\mathcal{T}$ is a two-dimensional convolution operation, image super-resolution when $\mathcal{T}$ is a composite operation of convolution and down-sampling, color image demosaicing when $\mathcal{T}$ is a color filter array (CFA) masking operation.

Since IR is an ill-posed inverse problem, the prior which is also called regularization needs to be adopted to constrain the solution space~\cite{roth2009fields,zoran2011learning}.
From a Bayesian perspective, the solution $\hat{\mathbf{x}}$ can be obtained by solving a Maximum A Posteriori (MAP) estimation problem,
\begin{equation}\label{eq1}
  \hat{\mathbf{x}} = \mathop{\arg\max}\limits_\mathbf{x} \log p(\mathbf{y}|\mathbf{x}) + \log p(\mathbf{x}),
\end{equation}
where $\log p(\mathbf{y}|\mathbf{x})$ represents the log-likelihood of observation $\mathbf{y}$, $\log p(\mathbf{x})$ delivers the prior of clean image $\mathbf{x}$ and is independent of degraded image $\mathbf{y}$.
More formally, \eqref{eq1} can be reformulated as
\begin{equation}\label{eq2}
  \hat{\mathbf{x}} = \mathop{\arg\min}_\mathbf{x}  \frac{1}{2\sigma^2}\|\mathbf{y} - \mathcal{T}(\mathbf{x})\|^2 + \lambda \mathcal{R}(\mathbf{x}),
\end{equation}
where the solution minimizes an energy function composed of a data term $\frac{1}{2\sigma^2}\|\mathbf{y} - \mathcal{T}(\mathbf{x})\|^2$ and a  regularization term $\lambda\mathcal{R}(\mathbf{x})$ with regularization parameter $\lambda$. Specifically, the data term guarantees the solution accords with the degradation process, while the prior term alleviates the ill-posedness by enforcing desired property on the solution.

Generally, the methods to solve \eqref{eq2} can be divided into two main categories, i.e., model-based methods and learning-based methods.
The former aim to directly solve \eqref{eq2} with some optimization algorithms, while the latter mostly train a truncated unfolding inference through an optimization of a loss function on a training set containing $N$ degraded-clean image pairs $\{(\mathbf{y}_i, \mathbf{x}_{i})\}_{i=1}^N$~\cite{tappen2007utilizing,barbu2009training,sun2013separable,schmidt2014shrinkage,chen2015trainable}. In particular, the learning-based methods are usually modeled as the following bi-level optimization problem
\begin{subequations}\label{eq_bilevel}
\begin{numcases}{}
\min_\Theta \sum_{i=1}^N\mathcal{L}(\mathbf{\hat{x}}_i, \mathbf{x}_i) \label{eq_bilevel_1}\\
s.t. \quad \hat{\mathbf{x}}_i = \mathop{\arg\min}_\mathbf{x}  \frac{1}{2\sigma^2}\|\mathbf{y}_i - \mathcal{T}(\mathbf{x})\|^2 + \lambda \mathcal{R}(\mathbf{x}), \label{eq_bilevel_2}
\end{numcases}
\end{subequations}
where $\Theta$ denotes the trainable parameters, $\mathcal{L}(\hat{\mathbf{x}}_i, \mathbf{x}_i)$ measures the loss of estimated clean image $\hat{\mathbf{x}}_i$ with respect to ground truth image $\mathbf{x}_i$. By replacing the unfolding inference~\eqref{eq_bilevel_2} with a predefined function $\mathbf{\hat{x}} = f(\mathbf{y}, \Theta)$, one can treat the plain learning-based methods as general case of~\eqref{eq_bilevel}.

It is easy to note that one main difference between model-based methods and learning-based methods is that, the former are flexible to handle various IR tasks by simply specifying $\mathcal{T}$ and can directly optimize on the degraded image $\mathbf{y}$,  {whereas the later require cumbersome training to learn the model before testing and are usually restricted by specialized tasks.}
Nevertheless, learning-based methods can not only enjoy a fast testing speed but also tend to deliver better performance due to the end-to-end training. In contrast, model-based methods are usually time-consuming with sophisticated priors for the purpose of good performance~\cite{gu2014weighted}.
As a result, these two categories of methods have their respective merits and drawbacks, and thus it would be attractive to investigate their integration which leverages their respective merits.
Such an integration has resulted in the deep plug-and-play IR method which replaces the denoising subproblem of model-based optimization with learning-based CNN denoiser prior.

The main idea of deep plug-and-play IR is that, with the aid of variable splitting algorithms, such as alternating direction method of multipliers (ADMM) ~\cite{boyd2011distributed} and half-quadratic splitting (HQS)~\cite{geman1995nonlinear}, it is possible to deal with the data term and prior term separately~\cite{parikh2014proximal}, and particularly, the prior term only corresponds to a denoising subproblem~\cite{danielyan2010image,heide2014flexisp,venkatakrishnan2013plug} which can be solved via deep CNN denoiser.
Although several deep plug-and-play IR works have been proposed, they typically suffer from the following drawbacks.
First, they either adopt different denoisers to cover a wide range of noise levels or use a single denoiser trained on a certain noise level, which are not suitable to solve the denoising subproblem.
For example, the IRCNN~\cite{zhang2017learning} denoisers involve 25 separate 7-layer denoisers, each of which is trained on an interval noise level of 2.
Second, their deep denoisers are not powerful enough, and thus, the performance limit of deep plug-and-play IR is unclear.
Third, a deep empirical understanding of their working mechanism is lacking.

This paper is an extension of our previous work~\cite{zhang2017learning} with a more flexible and powerful deep CNN denoiser which aims to push the limits of deep plug-and-play IR by conducting extensive experiments on different IR tasks. Specifically, inspired by FFDNet~\cite{zhang2018ffdnet}, the proposed deep denoiser can handle a wide range of noise levels via a single model by taking the noise level map as input. Moreover, its effectiveness is enhanced by taking advantages of both ResNet~\cite{he2015deep} and U-Net~\cite{ronneberger2015u}.
The deep denoiser is further incorporated into HQS-based plug-and-play IR to show the merits of using powerful deep denoiser. Meanwhile, a novel periodical geometric self-ensemble is proposed to potentially improve the performance without introducing extra computational burden, and a thorough analysis of parameter setting, intermediate results and empirical convergence are provided to better understand the working mechanism of the proposed deep plug-and-play IR.

The contribution of this work is summarized as follows:
\begin{itemize}
  \item A flexible and powerful deep CNN denoiser is trained. It not only outperforms the state-of-the-art deep Gaussian denoising models but also is suitable to solve plug-and-play IR.
  \item The HQS-based plug-and-play IR is thoroughly analyzed with respect to parameter setting, intermediate results and empirical convergence, providing a better understanding of the working mechanism.
  \item Extensive experimental results on deblurring, super-resolution and demosaicing have demonstrated the superiority of the proposed plug-and-play IR with deep denoiser prior.
\end{itemize}

\section{Related Works}

Plug-and-play IR generally involves two steps.
The first step is to decouple the data term and prior term of the objective function via a certain variable splitting algorithm, resulting in an iterative scheme consisting of alternately solving a data subproblem and a prior subproblem. The second step is to solve the prior subproblem with any off-the-shelf denoisers, such as K-SVD~\cite{elad2006image}, non-local means~\cite{buades2005non}, BM3D~\cite{dabov2007image}.
As a result, unlike traditional model-based methods which needs to specify the explicit and hand-crafted image priors, plug-and-play IR can implicitly define the prior via the denoiser. Such an advantage offers the possibility of leveraging very deep CNN denoiser to improve effectiveness.

\subsection{Plug-and-Play IR with Non-CNN Denoiser}
The plug-and-play IR can be traced back to~\cite{danielyan2010image,zoran2011learning,venkatakrishnan2013plug}.
In~\cite{danielyan2012bm3d}, Danielyan et al. used Nash equilibrium to derive an iterative decoupled deblurring BM3D (IDDBM3D) method for image debluring.   In~\cite{egiazarian2015single}, a similar method equipped with CBM3D denoiser prior was proposed for single image super-resolution (SISR). By iteratively updating a back-projection step and a CBM3D denoising step, the method has an encouraging performance for its PSNR improvement over SRCNN~\cite{dong2016}.
In~\cite{danielyan2010image}, the augmented Lagrangian method was adopted to fuse the BM3D denoiser to solve image deblurring task. With a similar iterative scheme as in~\cite{danielyan2012bm3d},
the first work that treats the denoiser as ``plug-and-play prior'' was proposed in~\cite{venkatakrishnan2013plug}.
Prior to that, a similar plug-and-play idea is mentioned in~\cite{zoran2011learning} where HQS algorithm is adopted for image denoising, deblurring and inpainting.
In~\cite{heide2014flexisp}, Heide et al. used an alternative to ADMM and HQS, i.e., the primal-dual algorithm~\cite{chambolle2011first}, to decouple the data term and prior term.
In~\cite{teodoro2016image}, Teodoro et al. plugged class-specific Gaussian mixture model (GMM) denoiser~\cite{zoran2011learning} into ADMM to solve image deblurring and compressive imaging.
In~\cite{metzler2016denoising}, Metzler et al. developed a denoising-based approximate message passing (AMP) method to integrate denoisers, such as BLS-GSM~\cite{portilla2003image} and BM3D, for compressed sensing reconstruction.
In~\cite{chan2016plug}, Chan et al. proposed plug-and-play ADMM algorithm with BM3D denoiser for single image super-resolution and quantized Poisson image recovery for single-photon imaging.
In~\cite{kamilov2017plug}, Kamilov et al. proposed fast iterative shrinkage thresholding algorithm (FISTA) with BM3D and WNNM~\cite{gu2014weighted} denoisers for nonlinear inverse scattering.
In~\cite{sun2019regularized}, Sun et al. proposed FISTA by plugging TV and BM3D denoiser prior for Fourier ptychographic microscopy.
In~\cite{yair2018multi}, Yair and Michaeli proposed to use WNNM denoiser as the plug-and-play prior for inpainting and deblurring.
In~\cite{gavaskar2020plug}, Gavaskar and Chaudhury investigated the convergence of ISTA-based plug-and-play IR with non-local means denoiser.

\begin{figure*}[!tbp]
\centering
\begin{overpic}[width=0.99\textwidth]{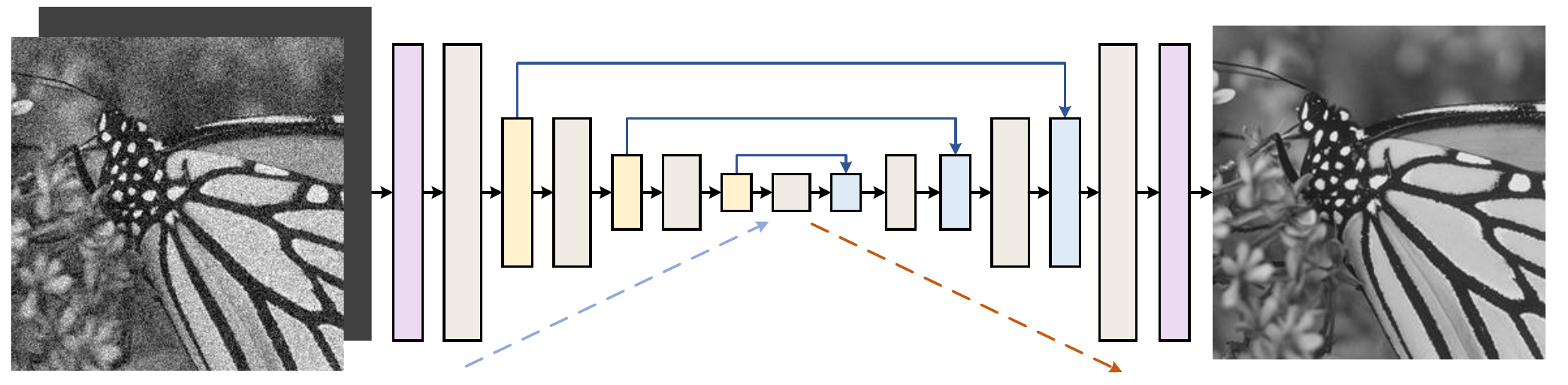}
%\put(10,4){\color{black}{\small Blur Kernel}}
\begin{turn}{90}
%\begin{sideways}
\put(10.2,-26.5){\color{black}{\small Conv}}
\put(5.2,-30){\color{black}{\small 4 Residual Blocks}}
\put(9.6,-33.5){\color{black}{\small SConv}}
\put(9.6,-68.5){\color{black}{\small TConv}}
\put(5.2,-71.8){\color{black}{\small 4 Residual Blocks}}
\put(10.2,-75.4){\color{black}{\small Conv}}
%\end{sideways}
\end{turn}
\put(-31,18.2){\color{black}{\small Skip Connection}}
\begin{turn}{26.2}
%\begin{sideways}
\put(-44,24.4){\color{black}{\small Downscaling}}
%\end{sideways}
\end{turn}
\begin{turn}{-26.2}
%\begin{sideways}
\put(-35.4,-10.1){\color{black}{\small Upscaling}}
%\end{sideways}
\end{turn}
\put(-85.8,20.7){\color{black}{\small \textcolor[rgb]{1.00,1.00,1.00}{Noisy Image}}}
\put(-85.8,22.6){\color{black}{\small \textcolor[rgb]{1.00,1.00,1.00}{Noise Level Map}}}
\put(-10.4,21.2){\color{black}{\small \textcolor[rgb]{1.00,1.00,1.00}{Denoised Image}}}
\end{overpic}
\caption{The architecture of the proposed DRUNet denoiser prior. DRUNet takes an additional noise level map as input and combines U-Net~\cite{ronneberger2015u} and ResNet~\cite{he2016deep}. ``SConv''  and ``TConv'' represent strided convolution and transposed convolution, respectively.}
\label{network}%\vspace{-0.15cm}
\end{figure*}
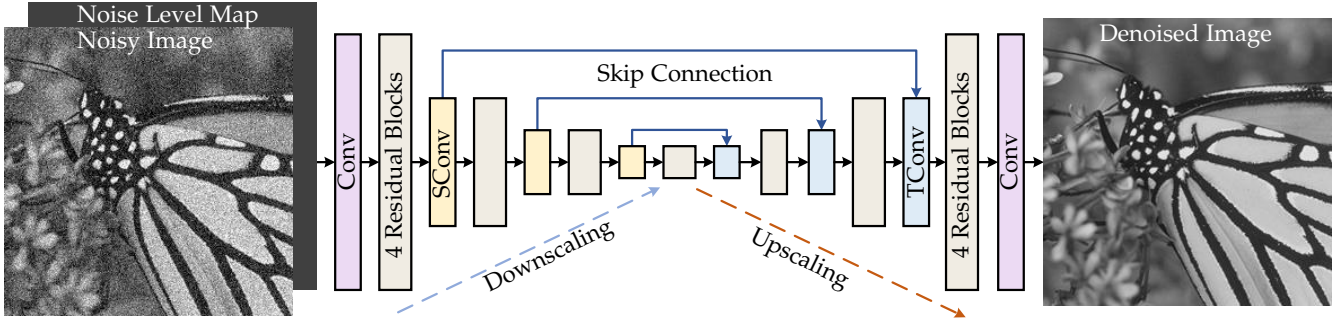

\subsection{Plug-and-Play IR with Deep CNN Denoiser}
With the development of deep learning techniques such as network design and gradient-based optimization algorithm, CNN-based denoiser has shown promising performance in terms of effectiveness and efficiency. Following its success, a flurry of CNN denoiser based plug-and-play IR works have been proposed.
In~\cite{romano2017little}, Romano et al. proposed explicit regularization by TNRD denoiser for image deblurring and SISR. In our previous work~\cite{zhang2017learning}, different CNN denoisers are trained to plug into HQS algorithm to solve deblurring and SISR. In~\cite{tirer2018image}, Tirer and Giryes proposed iterative denoising and backward projections with IRCNN denoisers for image inpainting and deblurring.
In~\cite{gu2018integrating}, Gu et al. proposed to adopt WNNM and IRCNN denoisers for plug-and-play deblurring and SISR.
In~\cite{tirer2019super}, Tirer and Giryes proposed use the IRCNN denoisers for plug-and-play SISR.
In~\cite{li2019learning}, Li and Wu plugged the IRCNN denoisers into the split Bregman iteration algorithm to solve depth image inpainting.
In~\cite{ryu2019plug}, Ryu et al. provided the theoretical convergence analysis of plug-and-play IR based on forward-backward splitting algorithm and ADMM algorithm, and proposed spectral normalization to train a DnCNN denoiser.
In~\cite{sun2019block}, Sun et al. proposed a block coordinate regularization-by-denoising (RED) algorithm by leveraging DnCNN~\cite{zhang2017beyond} denoiser as the explicit regularizer.

Although plug-and-play IR can leverage the powerful expressiveness of CNN denoiser, existing methods generally exploit DnCNN or IRCNN which do not make full use of CNN. Typically, the denoiser for plug-and-play IR should be non-blind and requires to handle a wide range of noise levels. However, DnCNN needs to separately learn a model for each noise level. To reduce the number of denoisers, some works adopt one denoiser fixed to a small noise level. However, according to~\cite{romano2017little} and as will be shown in Sec.~\ref{sec:parametersetting}, such a strategy tends to require a large number of iterations for a satisfying performance which would increase the computational burden.
While IRCNN denoisers can handle a wide range of noise levels, it consists of 25 separate 7-layer denoisers, among which each denoiser is trained on an interval noise level of 2. Such a denoiser suffers from two drawbacks. First, it does not have the flexibility to hand a specific noise level. Second, it is not effective enough due to the shallow layers.
Given the above considerations, it is necessary to devise a flexible and powerful denoiser to boost the performance of plug-and-play IR.

\subsection{ {Why not Use a Blind Gaussian Denoiser for Plug-and-Play IR?}}
 {It is worth to emphasize that the denoiser for plug-and-play IR should be designed for non-blind Gaussian denoising. The reason is two-fold. First, as will be shown in~\eqref{eq33_2} of the iterative solution for plug-and-play IR, the sub-problem actually corresponds to a non-blind Gaussian denoising problem with a Gaussian noise level. Second, although the non-blind Gaussian denoising problem could be solved via the blind Gaussian denoiser, the role of the Gaussian denoiser for plug-and-play IR is to smooth out the unknown noise (e.g., structural noise introduced during iterations) rather than remove the Gaussian noise. As will be shown in Sec.~\ref{discussion}, the noise distribution during iterations is usually non-Gaussian and varies across different IR tasks and even different iterations. Moreover, as we will see in Sec.~\ref{sec:blinddenoiser}, while the non-blind Gaussian denoiser can smooth out such non-Gaussian noise by setting a proper noise level, the blind Gaussian denoiser does not have such ability as it can only remove the Gaussian-like noise~\cite{zhang2018ffdnet}.}

\subsection{Difference to Deep Unfolding IR}
It should be noted that, apart from plug-and-play IR, deep unfolding IR~\cite{zhang2018ista,aggarwal2018modl,dong2018denoising,bertocchi2020deep} can also incorporate the advantages of both model-based methods and learning-based methods. The main difference between them is that the latter interprets a truncated unfolding optimization as an end-to-end trainable deep network and thus usually produce better results with fewer iterations.
However, deep unfolding IR needs separate training for each task. On the contrary, plug-and-play IR is easy to deploy without such additional training.

\begin{table*}[!tbp]\footnotesize
\caption{Average PSNR(dB) results of different methods with noise levels 15, 25 and 50 on the widely-used Set12 and BSD68~\cite{MartinFTM01,roth2009fields,zhang2017beyond} datasets. The best and second best results are highlighted in \textcolor[rgb]{1.00,0.00,0.00}{red} and \textcolor[rgb]{0.00,0.00,1.00}{blue} colors, respectively.} \vspace{-0.5cm}
\center
\hspace{0.05cm}
\begin{tabular}{p{1.1cm}<{\centering}p{0.95cm}<{\centering}|p{1.1cm}<{\centering}p{1.1cm}<{\centering}|p{1.1cm}<{\centering}p{1.1cm}<{\centering}p{1.1cm}<{\centering}p{1.1cm}<{\centering}p{1.1cm}<{\centering}|p{1.1cm}<{\centering}p{1.1cm}<{\centering}p{1.1cm}<{\centering}}
  \shline
 \multirow{2}{*}{Datasets} & Noise & \multirow{2}{*}{BM3D} & \multirow{2}{*}{WNNM} & \multirow{2}{*}{DnCNN}  & \multirow{2}{*}{$\text{N}^3$Net}  & \multirow{2}{*}{NLRN}& \multirow{2}{*}{RNAN}& \multirow{2}{*}{FOCNet} & \multirow{2}{*}{IRCNN}& \multirow{2}{*}{FFDNet} & \multirow{2}{*}{DRUNet}\\%\cline{4-10}

 & Level &  &   & &   & &  &  &  &    &    \\ \hline\hline
  & 15 & 32.37 &  32.70 &32.86   & --  & \textcolor{blue}{33.16} & -- & 33.07 &32.77 & 32.75  & \textcolor{red}{33.25}\\
 Set12 & 25 & 29.97 &  30.28 &30.44   & 30.55& \textcolor{blue}{30.80} & --  & 30.73 & 30.38 & 30.43 &\textcolor{red}{30.94}\\
  & 50 & 26.72 &  27.05  &27.18    & 27.43 & 27.64 & \textcolor{blue}{27.70} & 27.68 & 27.14 & 27.32 &\textcolor{red}{27.90}\\\hline

  & 15 & 31.08 &  31.37 &31.73   & --  & \textcolor{blue}{31.88}& -- & 31.83 &31.63& 31.63  & \textcolor{red}{31.91}\\
 BSD68 & 25 & 28.57 &  28.83 &29.23   & 29.30& \textcolor{blue}{29.41}& --  & 29.38 & 29.15 & 29.19 &\textcolor{red}{29.48}\\
  & 50 & 25.60 &  25.87  &26.23    & 26.39 & 26.47& 26.48 & \textcolor{blue}{26.50} & 26.19 & 26.29 &\textcolor{red}{26.59}\\\shline
\end{tabular}
\label{table_denoising_grayscale}%\vspace{-0.1cm}
\end{table*}

\section{Learning Deep CNN Denoiser Prior}
Although various CNN-based denoising methods have been recently proposed, most of them are not designed for plug-and-play IR.
In~\cite{lehtinen2018noise2noise,krull2019noise2void,batson2019noise2self}, a novel training strategy without ground-truth is proposed. In~\cite{guo2019toward,brooks2019unprocessing,abdelhamed2019noise,zamir2020cycleisp}, real noise synthesis technique is proposed to handle real digital photographs.
However, from a Bayesian perspective, the denoiser for plug-and-play IR should be a Gaussian denoiser. Hence, one can add synthetic Gaussian noise to clean image for supervised training.
In~\cite{lefkimmiatis2017non,zhang2019residual,liu2018non,plotz2018neural}, the non-local module was incorporated into the network design for better restoration. However, these methods learn a separate model for each noise level.
Perhaps the most suitable denoiser for plug-and-play IR is FFDNet~\cite{zhang2018ffdnet} which can handle a wide range of noise levels by taking the noise level map as input. Nevertheless, FFDNet only has a comparable performance to DnCNN and IRCNN, thus lacking effectiveness to boost the performance of plug-and-play IR. For this reason, we propose to improve FFDNet by taking advantage of the widely-used U-Net~\cite{ronneberger2015u} and ResNet~\cite{he2015deep} for architecture design.

\subsection{Denoising Network Architecture}

It is well-known that U-Net~\cite{ronneberger2015u} is effective and efficient for image-to-image translation, while ResNet~\cite{he2015deep} is superior in increasing the modeling capacity by stacking multiple residual blocks. Following FFDNet~\cite{zhang2018ffdnet} that takes the noise level map as input, the proposed denoiser, namely DRUNet, further integrates residual blocks into U-Net for effective denoiser prior modeling.  {Note that this work focuses on providing a flexible and powerful pre-trained denoiser to benefit existing plug-and-play IR methods rather than designing new denoising network architecture. Actually, the similar idea of combining U-Net and ResNet can also be found in other works such as~\cite{zhang2018road,venkatesh2018deep}.}

The architecture of DRUNet is illustrated in Fig.~\ref{network}. Like FFDNet, DRUNet has the ability to handle various noise levels via a single model. The backbone of DRUNet is U-Net which consists of four scales. Each scale has an identity skip connection between $2\times2$ strided convolution (SConv) downscaling and $2\times2$ transposed convolution (TConv) upscaling operations. The number of channels in each layer from the first scale to the fourth scale are $64$, $128$, $256$ and $512$, respectively.
Four successive residual blocks are adopted in the downscaling and upscaling of each scale. Inspired by the network architecture design for super-resolution in~\cite{lim2017enhanced}, no activation function is followed by the first and the last convolutional (Conv) layers, as well as SConv and TConv layers. In addition, each residual block only contains one ReLU activation function.

It is worth noting that the proposed DRUNet is bias-free, which means no bias is used in all the Conv, SConv and TConv layers. The reason is two-fold. First, bias-free network with ReLU activation and identity skip connection naturally enforces scaling invariance property of many image restoration tasks, i.e., $f(ax)=af(x)$ holds true for any scalar $a\geq0$ (please refer to~\cite{2019arXiv190605478M} for more details). Second, we have empirically observed that, for the network with bias, the magnitude of bias would be much larger than that of filters, which in turn may harm the generalizability.

\begin{figure*}[!htbp]
\begin{center}
\subfigure[Noisy (14.78dB)]{
\begin{overpic}[width=0.157\textwidth]{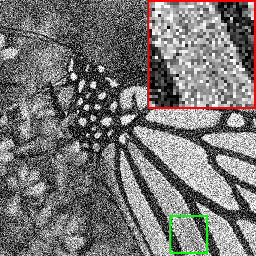}
\end{overpic}}
\subfigure[BM3D (25.82dB)]{
\begin{overpic}[width=0.157\textwidth]{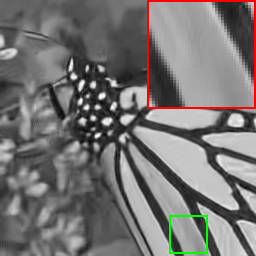}
\end{overpic}}
\subfigure[DnCNN (26.83dB)]{
\begin{overpic}[width=0.157\textwidth]{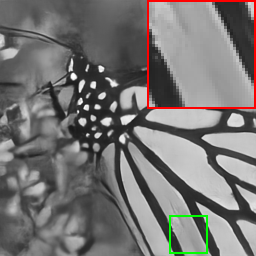}
\end{overpic}}
\subfigure[RNAN (27.18dB)]{
\begin{overpic}[width=0.157\textwidth]{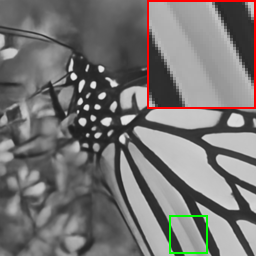}
\end{overpic}}
\subfigure[FFDNet (26.92dB)]{
\begin{overpic}[width=0.157\textwidth]{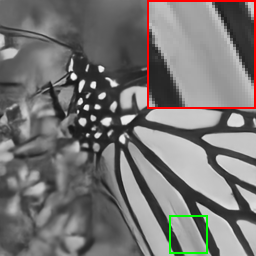}
\end{overpic}}
\subfigure[DRUNet (27.31dB)]{
\begin{overpic}[width=0.157\textwidth]{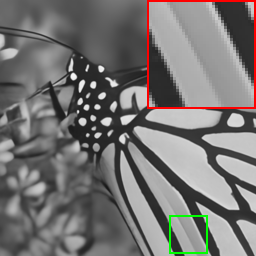}
\end{overpic}}
\vspace{-0.4cm}
\caption{Grayscale image denoising results of different methods on image ``\emph{Monarch}'' from Set12 dataset with noise level 50.}\label{fig:denoise_1}
\end{center}\vspace{-0.1cm}
\end{figure*}

\begin{figure*}[!htbp]
\begin{center}
\subfigure[Noisy (14.99dB)]{
\begin{overpic}[trim=2cm 0cm 0cm 0cm, clip=true,width=0.157\textwidth]{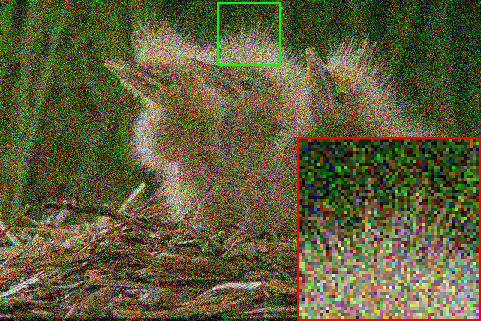}
\end{overpic}}
\subfigure[BM3D (28.36dB)]{
\begin{overpic}[trim=2cm 0cm 0cm 0cm, clip=true,width=0.157\textwidth]{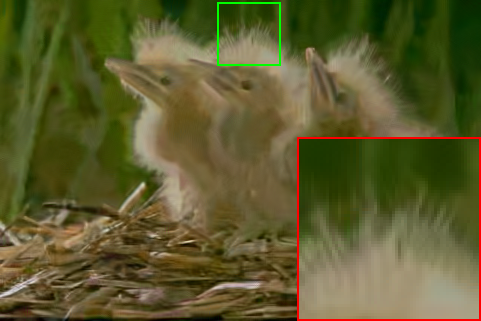}
\end{overpic}}
\subfigure[DnCNN (28.68dB)]{
\begin{overpic}[trim=2cm 0cm 0cm 0cm, clip=true,width=0.157\textwidth]{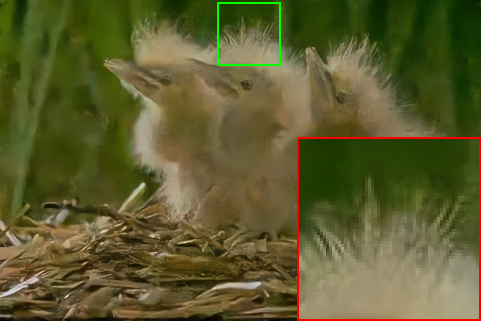}
\end{overpic}}
\subfigure[FFDNet (28.75dB)]{
\begin{overpic}[trim=2cm 0cm 0cm 0cm, clip=true,width=0.157\textwidth]{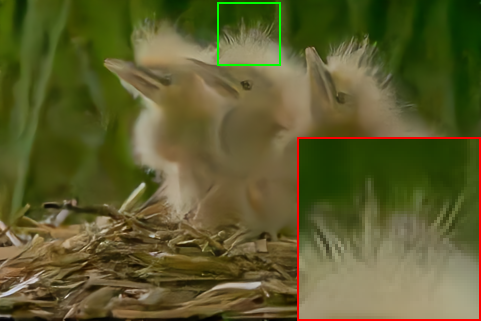}
\end{overpic}}
\subfigure[IRCNN (28.69dB)]{
\begin{overpic}[trim=2cm 0cm 0cm 0cm, clip=true,width=0.157\textwidth]{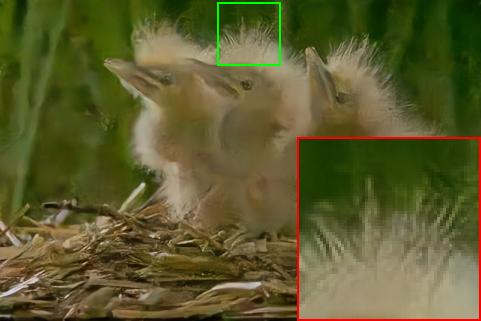}
\end{overpic}}
\subfigure[DRUNet (29.28dB)]{
\begin{overpic}[trim=2cm 0cm 0cm 0cm, clip=true,width=0.157\textwidth]{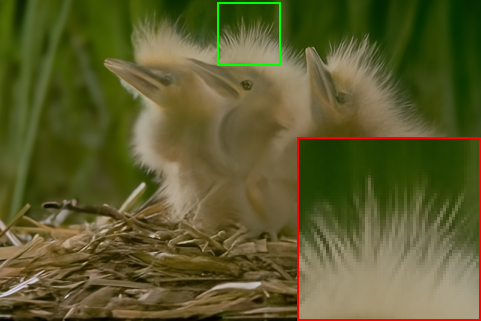}
\end{overpic}}
\vspace{-0.4cm}
\caption{Color image denoising results of different methods on image ``\emph{163085}'' from CBSD68 dataset with noise level 50.}\label{fig:denoise_2}
\end{center}\vspace{-0.1cm}
\end{figure*}

\subsection{Training Details}

It is well known that CNN benefits from the availability of large-scale training data. To enrich the denoiser prior for plug-and-play IR, instead of training on a small dataset that includes 400 Berkeley segmentation dataset (BSD) images of size 180$\times$180~\cite{chen2015trainable}, we construct a large dataset consisting of 400 BSD images, 4,744 images of Waterloo Exploration Database~\cite{ma2016gmad}, 900 images from DIV2K dataset~\cite{agustsson2017ntire}, and 2,750 images from Flick2K dataset~\cite{lim2017enhanced}.
Because such a dataset covers a larger image space, the learned model can slightly improve the PSNR results on BSD68 dataset~\cite{roth2009fields} while having an obvious PSNR gain on testing datasets from a different domain.

As a common setting for Gaussian denoising, the noisy counterpart $\mathbf{y}$ of clean image $\mathbf{x}$ is obtained by adding AWGN with noise level $\sigma$. Correspondingly, the noise level map is a uniform map filled with $\sigma$ and has the same spatial size as noisy image.
To handle a wide range of noise levels, the noise level $\sigma$ is randomly chosen from $[0, 50]$ during training.
 {Note that the noisy images are not clipped into the range of $[0, 255]$. The reason is that the clipping operation would change the distribution of the noise, which in turn will give rise to inaccurate solution for plug-and-play IR.}
The network parameters are optimized by minimizing the L1 loss rather than L2 loss between the denoised image and its ground-truth with Adam algorithm~\cite{kingma2014adam}.
Although there is no direct evidence on which loss would result in better performance, it is widely acknowledged that L1 loss is more robust than L2 loss in handling outliers~\cite{bishop2006pattern}. Regarding to denoising, outliers may occur during the sampling of AWGN. In this sense, L1 loss tends to be more stable than L2 loss for denoising network training.
The learning rate starts from 1e-4 and then decreases by half every 100,000 iterations and finally ends once it is smaller than 5e-7. In each iteration during training, 16 patches with patch size of 128$\times$128 were randomly sampled from the training data.
We separately learn a denoiser model for grayscale image and color image.
It takes about four days to train the model with PyTorch and an Nvidia Titan Xp GPU.

\subsection{Denoising Results}

\subsubsection{Grayscale Image Denoising}

For grayscale image denoising, we compared the proposed DRUNet denoiser with several state-of-the-art denoising methods, including two representative model-based methods (i.e., BM3D~\cite{dabov2007image} and WNNM~\cite{gu2014weighted}), five CNN-based methods which separately learn a single model for each noise level (i.e., DnCNN~\cite{zhang2017beyond}, $\text{N}^3$Net~\cite{plotz2018neural}, NLRN~\cite{liu2018non}, RNAN~\cite{zhang2019rnan}, FOCNet~\cite{jia2019focnet}) and two CNN-based methods which were trained to handle a wide range of noise levels (i.e., IRCNN~\cite{zhang2017learning} and FFDNet~\cite{zhang2018ffdnet}).
Note that $\text{N}^3$Net, NLRN and RNAN adopt non-local module in the network architecture design so as to exploit non-local image prior. The PSNR results of different methods on the widely-used Set12~\cite{zhang2017beyond} and BSD68~\cite{MartinFTM01,roth2009fields} datasets for noise levels 15, 25 and 50 are reported in Table~\ref{table_denoising_grayscale}.
It can be seen that DRUNet achieves the best PSNR results for all the noise levels on the two datasets. Specifically, DRUNet has an average PSNR gain of about 0.9dB over BM3D and surpasses DnCNN, IRCNN and FFDNet by an average PSNR of 0.5dB on Set12 dataset and 0.25dB on BSD68 dataset. Despite the fact that NLRN, RNAN and FOCNet learn a separate model for each noise level and have a very competitive performance, they fail to outperform DRUNet.
Fig.~\ref{fig:denoise_1} shows the grayscale image denoising results of different methods on image ``\emph{Monarch}'' from Set12 dataset with noise level 50. It can be seen that DRUNet can recover much sharper edges than BM3D, DnCNN, FFDNet while having similar result with RNAN.

\subsubsection{Color Image Denoising}

Since existing methods mainly focus on grayscale image denoising, we only compare DRUNet with CBM3D, DnCNN, IRCNN and FFDNet for color denoising.
Table~\ref{table_denoising_color} reports the color image denoising results of different methods for noise levels 15, 25 and 50 on CBSD68~\cite{MartinFTM01,roth2009fields,zhang2017beyond}, Kodak24~\cite{franzen1999kodak} and McMaster~\cite{zhang2011color} datasets. One can see that DRUNet outperforms the other competing methods by a large margin.
It is worth noting that while having a good performance on CBSD68 dataset, DnCNN does not perform well on McMaster dataset. Such a discrepancy highlights the importance of reducing the image domain gap between training and testing for image denoising.
The visual results of different methods on image ``\emph{163085}'' from CBSD68 dataset with noise level 50 are shown in Fig.~\ref{fig:denoise_2} from which it can be seen that DRUNet can recover more fine details and textures than the competing methods.

\begin{table}[!htbp]\footnotesize
\caption{Average PSNR(dB) results of different methods for noise levels 15, 25 and 50 on CBSD68~\cite{MartinFTM01,roth2009fields,zhang2017beyond}, Kodak24 and McMaster datasets. The best and second best results are highlighted in \textcolor[rgb]{1.00,0.00,0.00}{red} and \textcolor[rgb]{0.00,0.00,1.00}{blue} colors, respectively.} %\vspace{-0.25cm}
\center
\begin{tabular}{p{0.8cm}<{\centering}p{0.8cm}<{\centering}|p{0.8cm}<{\centering}p{0.8cm}<{\centering}p{0.8cm}<{\centering}p{0.85cm}<{\centering}p{0.9cm}<{\centering}}
  \shline
 \multirow{2}{*}{Datasets} & Noise & \multirow{2}{*}{CBM3D} & \multirow{2}{*}{DnCNN} & \multirow{2}{*}{IRCNN}& \multirow{2}{*}{FFDNet} & \multirow{2}{*}{DRUNet}\\%\cline{4-10}

 & Level & & &  &  &       \\ \hline\hline
  & 15 & 33.52 & \textcolor[rgb]{0.00,0.00,1.00}{33.90} &  33.86 &33.87   & \textcolor{red}{34.30}  \\
 CBSD68 & 25 & 30.71 & \textcolor[rgb]{0.00,0.00,1.00}{31.24} & 31.16 &31.21  & \textcolor{red}{31.69}\\
  & 50 & 27.38 & 27.95 & 27.86  &\textcolor[rgb]{0.00,0.00,1.00}{27.96}   & \textcolor{red}{28.51} \\\hline

  & 15 & 34.28 & 34.60 &\textcolor[rgb]{0.00,0.00,1.00}{34.69} &34.63   & \textcolor{red}{35.31}  \\
 Kodak24 & 25 & 32.15& 32.14 &  \textcolor[rgb]{0.00,0.00,1.00}{32.18} &32.13   & \textcolor{red}{32.89}\\
  & 50 & 28.46 & 28.95 &28.93  &\textcolor[rgb]{0.00,0.00,1.00}{28.98}    & \textcolor{red}{29.86} \\\hline

  & 15 & 34.06 & 33.45 &34.58 &\textcolor[rgb]{0.00,0.00,1.00}{34.66}   & \textcolor{red}{35.40}  \\
 McMaster & 25 &31.66& 31.52  &  32.18 &\textcolor[rgb]{0.00,0.00,1.00}{32.35}   & \textcolor{red}{33.14}\\
  & 50 & 28.51 & 28.62 &28.91  &\textcolor[rgb]{0.00,0.00,1.00}{29.18}    & \textcolor{red}{30.08} \\\shline

\end{tabular}
\label{table_denoising_color}%\vspace{-0.1cm}
\end{table}

\subsubsection{Extended Application to JPEG Image Deblocking}

\begin{table}[!bp]\footnotesize
\caption{ {Average PSNR(dB) results of different methods for JPEG image deblocking with quality factors 10, 20, 30 and 40 on Classic5 and LIVE1 datasets.} The best and second best results are highlighted in \textcolor[rgb]{1.00,0.00,0.00}{red} and \textcolor[rgb]{0.00,0.00,1.00}{blue} colors, respectively.} \vspace{-0.25cm}
\center
\begin{tabular}{p{0.7cm}<{\centering}p{0.4cm}<{\centering}|p{0.7cm}<{\centering}p{0.6cm}<{\centering}p{0.7cm}<{\centering}p{0.65cm}<{\centering}p{0.60cm}<{\centering}p{0.8cm}<{\centering}}
  \shline
 \multirow{2}{*}{Datasets} & \multirow{2}{*}{$q$} & \multirow{2}{*}{\scriptsize ARCNN} & \multirow{2}{*}{\scriptsize TNRD} & \multirow{2}{*}{\scriptsize DnCNN3}& \multirow{2}{*}{\scriptsize RNAN}& \multirow{2}{*}{\scriptsize QGAC} & \multirow{2}{*}{\scriptsize DRUNet}\\%\cline{4-10}

 &  & & &  &  &       \\ \hline\hline
\multirow{4}{*}{Classic5}  & 10 & 29.03 & 29.28 &  29.40 &\textcolor[rgb]{0.00,0.00,1.00}{29.96} & 29.84 & \textcolor{red}{30.16}  \\
 & 20 & 31.15 & 31.47 & 31.63 &\textcolor[rgb]{0.00,0.00,1.00}{32.11}& 31.98 & \textcolor{red}{32.39}\\
  & 30 & 32.51 & 32.78 & 32.91  &\textcolor[rgb]{0.00,0.00,1.00}{33.38}& 33.22  & \textcolor{red}{33.59} \\
  & 40 & 33.34 & - & 33.77  &\textcolor[rgb]{0.00,0.00,1.00}{34.27}& 34.05  & \textcolor{red}{34.41} \\\hline

\multirow{4}{*}{LIVE1}  & 10 & 28.96 & 29.15 &  29.19 &\textcolor[rgb]{0.00,0.00,1.00}{29.63}& 29.51 & \textcolor{red}{29.79}  \\
 & 20 & 31.29 & 31.46 & 31.59 &\textcolor[rgb]{0.00,0.00,1.00}{32.03}& 31.83 & \textcolor{red}{32.17}\\
  & 30 & 32.67 & 32.84 & 32.98 &\textcolor[rgb]{0.00,0.00,1.00}{33.45}& 33.20  & \textcolor{red}{33.59} \\
  & 40 & 33.63 & - & 33.96  &\textcolor[rgb]{0.00,0.00,1.00}{34.47}& 34.16  & \textcolor{red}{34.58} \\\shline

\end{tabular}
\label{table_deblocking}
\end{table}

The proposed DRUNet is also applicable to remove other different noise types, such as JPEG compression artifacts. 
By simply changing the training data and replacing the noise level $\sigma$ of AWGN with quality factor $q$ of JPEG compression, a DRUNet model for JPEG image deblocking is trained. We set the quality factor range to $[10, 95]$, where quality factor 10 represents lower quality and higher compression, and vice versa. Specifically, the quality factor $q$ are normalized with $(100-q)/100$ for 0-1 normalization. 
We use the same training data as in denoising for training.
Table~\ref{table_deblocking} reports the average PSNR(dB) results of different methods for JPEG image deblocking with quality factors 10, 20, 30 and 40 on Classic5 and LIVE1 datasets. The compared methods include ARCNN~\cite{dong2015compression}, TNRD~\cite{chen2015trainable}, DnCNN3~\cite{zhang2017beyond} and RNAN~\cite{zhang2019residual}, QGAC~\cite{ehrlich2020quantization}. Since DnCNN3 is trained also for denosing and SISR, we re-trained a non-blind DnCNN3 model with our training data.
Compared to the original DnCNN3 model, the new model has average PSNR gains of 0.21dB and 0.19dB on the two testing datasets. To quantify the performance contribution of training data, we also trained a DRUNet model with less training  data as in~\cite{zhang2018ffdnet}. The results show that the PSNR decreases by 0.04dB on average, which demonstrates that a large training data can slightly improve the performance for JPEG image deblocking.
From Table~\ref{table_deblocking}, we can see that DRUNet outperforms ARCNN, TNRD, DnCNN3 and QGAC by a large margin and has an average PSNR gain of 0.15dB over RNAN, which further demonstrates the flexibility and effectiveness of the proposed DRUNet.

\subsubsection{Generalizability to Unseen Noise Level}
\begin{figure}[!htbp]
\scriptsize{
\begin{center}
\subfigure[Noisy (7.50dB)]
{\includegraphics[trim=0cm 3cm 0cm 0cm, clip=true, width=0.156\textwidth]{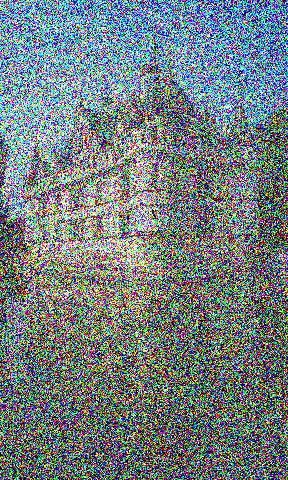}}\hspace{0.02cm}
\subfigure[DnCNN (22.16dB)]
{\includegraphics[trim=0cm 3cm 0cm 0cm, clip=true,width=0.156\textwidth]{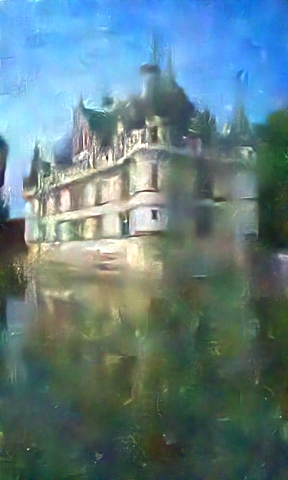}}\hspace{0.02cm}
\subfigure[IRCNN (22.22dB)]
{\includegraphics[trim=0cm 3cm 0cm 0cm, clip=true,width=0.156\textwidth]{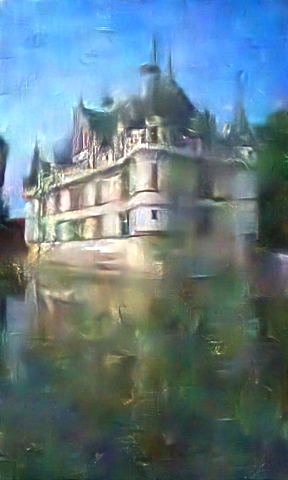}}\hspace{0.0001cm}
\subfigure[FFDNet (20.97dB)]
{\includegraphics[trim=0cm 3cm 0cm 0cm, clip=true,width=0.156\textwidth]{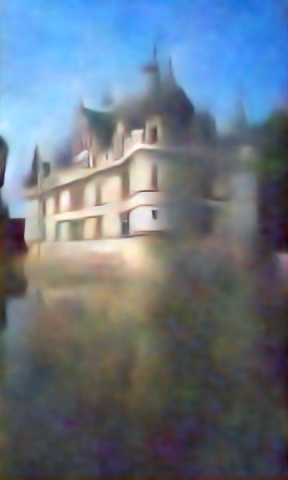}}\hspace{0.02cm}
\subfigure[\scriptsize{DRUNet+B} (18.46dB)]
{\includegraphics[trim=0cm 3cm 0cm 0cm, clip=true,width=0.156\textwidth]{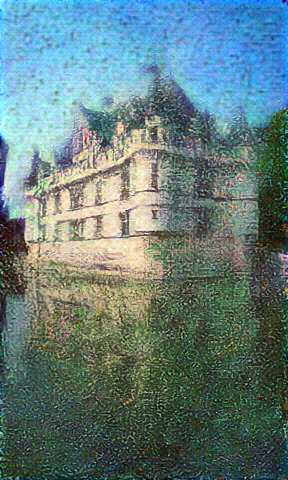}}\hspace{0.02cm}
\subfigure[DRUNet (23.55dB)]
{\includegraphics[trim=0cm 3cm 0cm 0cm, clip=true,width=0.156\textwidth]{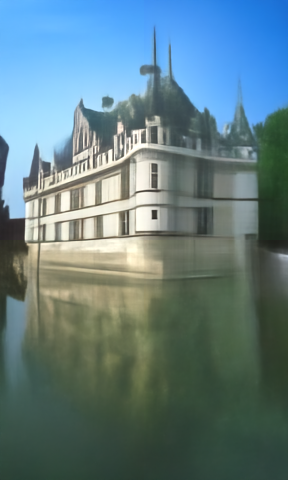}}
\vspace{-0.2cm}
\caption{ {An example to show the generalizability advantage of the proposed bias-free DRUNet. The noise level of the noisy image is 200.}}\label{fig:denoise_3}
\end{center}}\vspace{-0.2cm}
\end{figure}

In order to show the advantage of the bias-free DRUNet, we also train a DRUNet+B model whose biases were randomly initialized from a uniform distribution in $[-1, 1]$.
Fig.~\ref{fig:denoise_3} provides the visual results comparison between different models on a noisy image with an extremely large unseen noise level of 200. 
Note that since DnCNN and IRCNN do not have the flexibility to change the noise level, we first multiply the noisy image by a factor of 0.25 so that the noise level changes from 200 to 50. We then apply the DnCNN and IRCNN models for denoising and finally obtain the denoising results with a multiplication of 4.
From Fig.~\ref{fig:denoise_3}, we can seen that, even trained on noise level range of $[0, 50]$, the bias-free DRUNet can still perform well, whereas DRUNet+B (with biases) introduces noticeable visual artifacts while having a much lower PSNR. As we will see in Sec.~\ref{ablation}, DRUNet+B has a comparable performance with bias-free DRUNet for noise levels in $[0, 50]$.
Thus, we can conclude that bias-free DRUNet can enhance the generalizability to unseen noise level. Note that whether bias-free network can benefit other tasks or not remains further study.

\subsubsection{Ablation Study}\label{ablation}
 {In order to further analyze the proposed DRUNet, we have performed an ablation study to quantify the performance contribution of different factors such as residual blocks, training data, biases, and noise level map. Table~\ref{table:ablation} reports the comparisons between DRUNet with four different cases, including case 1: DRUNet without skip connections of the residual blocks, case 2: DRUNet with less training data as in~\cite{zhang2018ffdnet}, case 3: DRUNet without removing the biases (i.e., DRUNet+B), and case 4: DRUNet without taking noise level map as input. By comparison, we can have the following conclusions. First, the residual blocks can ease the training to get a better performance. Second, the performance on Set12 dataset tends to be saturated with enough training data as the DRUNet model with more training data only improves the PSNR by an average of 0.01dB. Third, DRUNet with biases has a similar performance to bias-free DRUNet for the trained noise levels, however, the bias-free DRUNet can enhance the generalizability to unseen noise level. Fourth, the noise level map can improve the performance as it introduces extra information of the noise. Such a phenomenon has also been reported in~\cite{zhang2018ffdnet}.}

\begin{table}[!htbp]\footnotesize
\caption{ {The PSNR results comparison with different cases (i.e, case 1: DRUNet without residual blocks, case 2: DRUNet with less training data as in~\cite{zhang2018ffdnet}, case 3: DRUNet without removing the biases, and case 4: DRUNet without taking noise level map as input) with noise levels 15, 25 and 50 on Set12 dataset.}}\vspace{-0.3cm}
\center
\begin{tabular}{p{0.8cm}<{\centering}p{1.1cm}<{\centering}|p{0.7cm}<{\centering}p{0.7cm}<{\centering}p{0.8cm}<{\centering}p{0.7cm}<{\centering}|p{1.0cm}<{\centering}}
 \shline
\multirow{2}{*}{Datasets} &\multirow{1}{*}{Noise} &\multirow{2}{*}{Case 1} & \multirow{2}{*}{Case 2}  & \multirow{2}{*}{Case 3} & \multirow{2}{*}{Case 4} & \multirow{2}{*}{DRUNet} \\
& Level & & & & &\\\hline\hline

\multirow{3}{*}{Set12}&  15  &  33.12 & 33.23  & 33.24& 33.16 & 33.25  \\
&  25  & 30.82 & 30.92  & 30.93&30.86 &  30.94 \\
&  50  & 27.78 & 27.87  & 27.89 &27.83&  27.90 \\
\shline
\end{tabular}
\label{table:ablation}
\end{table}

\subsubsection{Runtime, FLOPs and Maximum GPU Memory Consumption}
 {Table~\ref{table:runtime} reports the runtime, FLOPs and maximum GPU memory consumption comparison with four representative methods (i.e., DnCNN, IRCNN, FFDNet and RNAN) on images of size 256$\times$256 and 512$\times$512 with noise level 50. We do not report other methods such as FOCNet for comparison since they are not implemented in PyTorch for a fair and easy comparison. 
Note that, in order to reduce the memory caused by the non-local module, RNAN splits the input image into overlapped patches with predefined maximum spatial size and then aggravates the results to obtain the final denoised image. The default maximum spatial size is 10,000 which is equivalent to a size of 100$\times$100. We also compare $\text{RNAN}^*$ which sets maximum spatial size to 70,000. As a simple example, RNAN and $\text{RNAN}^*$ splits an image of size 512$\times$512 into 64 and 4 overlapped patches, respectively. It should be noted that NLRN which also adopts a similar non-local module as RNAN uses a different strategy reduce the memory, i.e, fixing the patch size to 43$\times$43. However, it uses a small stride of 7 which would largely increase the computational burden.}

 {From Table~\ref{table:runtime}, one can see that FFDNet achieves the best performance on runtime, FLOPs and memory. Compared to DnCNN, DRUNet has much better PSNR values and only doubles the runtime, 
triples the FLOPs, and quadruples the maximum GPU memory consumption. In contrast, RNAN is about 60 times slower than DnCNN and also has a much larger FLOPs. In addition, it would dramatically aggravate the maximum GPU memory consumption with the increase of the predefined maximum spatial size. 
Note that RNAN does not outperform DRUNet in terms of PSNR. Such a phenomenon highlights that the non-local module in RNAN may not be a proper way to improve PSNR and further study is required to improve the runtime, FLOPs and maximum GPU memory consumption. }

\begin{table}[!tbp]\footnotesize
\caption{ {Runtime (in seconds), FLOPs (in G) and max GPU memory (in GB)  of different methods on images of size 256$\times$256 and 512$\times$512 with noise level 50. The experiments are conducted in PyTorch on a PC with
an Intel Xeon 3.5GHz 4-core CPU, 4-8GB of RAM and an Nvidia Titan Xp GPU.}}\vspace{-0.3cm}
\center
\begin{tabular}{p{0.8cm}<{\centering}p{1.0cm}<{\centering}|p{.55cm}<{\centering}p{0.55cm}<{\centering}p{0.6cm}<{\centering}p{0.6cm}<{\centering}p{0.6cm}<{\centering}p{0.7cm}<{\centering}}
 \shline
\multirow{2}{*}{Metric} &\multirow{1}{*}{Image} &\multirow{2}{*}{\scriptsize DnCNN}&\multirow{2}{*}{\scriptsize IRCNN}&\multirow{2}{*}{\scriptsize FFDNet} & \multirow{2}{*}{\scriptsize RNAN}  & \multirow{2}{*}{\scriptsize $\text{RNAN}^*$} & \multirow{2}{*}{\scriptsize DRUNet} \\
& Size & & & & \\\hline\hline
\multirow{2}{*}{Runtime}& 256$\times$256   & 0.0087 & 0.0066  & 0.0023  & 0.4675& 0.4662  & 0.0221  \\
& 512$\times$512    & 0.0314 & 0.0257  & 0.0071   & 2.1530& 1.8769   & 0.0733  \\\hline
\multirow{2}{*}{FLOPs}& 256$\times$256 & 36.38 & 12.18 &  7.95  & 774.67 & 495.79  & 102.91 \\
& 512$\times$512  & 145.52 &  48.72  & 31.80  & 3416.29  & 2509.92 & 411.65 \\\hline
\multirow{2}{*}{Memory}& 256$\times$256   & 0.0339 & 0.0346  & 0.0114  &  0.3525 & 2.9373 & 0.2143 \\
& 512$\times$512    & 0.1284 & 0.1360 & 0.0385  & 0.4240 & 3.2826  & 0.4911 \\

\shline
\end{tabular}
\label{table:runtime}
\end{table}

According to the above results, we can conclude that DRUNet is a flexible and powerful denoiser prior for plug-and-play IR.

\section{HQS Algorithm for Plug-and-play IR}

Although there exist various variable splitting algorithms for plug-and-play IR, HQS owes its popularity to the simplicity and fast convergence. We therefore adopt HQS in this paper. On the other hand, there is no doubt that parameter setting is always a non-trivial issue~\cite{romano2017little}. In other words, careful parameter setting is needed to obtain a good performance. To have a better understanding on the HQS-based plug-and-play IR, we will discuss the general methodology for parameter setting after providing the HQS algorithm. We then propose a periodical geometric self-ensemble strategy to potentially improve the performance.

\subsection{Half Quadratic Splitting (HQS) Algorithm}

In order to decouple the data term and prior term of~\eqref{eq2},
HQS first introduces an auxiliary variable $\mathbf{z}$, resulting in a constrained optimization problem given by
\begin{equation}\label{eq:constrained}
  \hat{\mathbf{x}} = \mathop{\arg\min}_\mathbf{x} ~ \frac{1}{2\sigma^2}\|\mathbf{y} - \mathcal{T}(\mathbf{x})\|^2 + \lambda \mathcal{R}(\mathbf{z})  \quad s.t. \quad  \mathbf{z} = \mathbf{x}.
\end{equation}
\eqref{eq:constrained} is then solved by minimizing the following problem
\begin{equation}\label{eq22}
  \mathcal{L}_\mu(\mathbf{x},\mathbf{z}) = \frac{1}{2\sigma^2}\|\mathbf{y} - \mathcal{T}(\mathbf{x})\|^2 + \lambda \mathcal{R}(\mathbf{z}) +  \frac{\mu}{2}\|\mathbf{z}-\mathbf{x}\|^2,
\end{equation}
where $\mu$ is a penalty parameter.
Such problem can be addressed by iteratively solving the following subproblems for $\mathbf{x}$ and $\mathbf{z}$ while keeping the rest of the variables fixed,
\begin{subequations}\label{eq33}
\begin{numcases}{}
\mathbf{x}_{k}=\mathop{\arg\min}_{\mathbf{x}} \|\mathbf{y} - \mathcal{T}(\mathbf{x})\|^2 + \mu\sigma^2\|\mathbf{x}-\mathbf{z}_{k-1} \|^2 \label{eq33_1}\\
\mathbf{z}_{k}=\mathop{\arg\min}_{\mathbf{z}} \frac{1}{2(\sqrt{\lambda/\mu})^2}\|\mathbf{z}-\mathbf{x}_{k}\|^2  + \mathcal{R}(\mathbf{z}). \label{eq33_2}
\end{numcases}
\end{subequations}
As such, the data term and prior term are decoupled into two separate subproblems. To be specific, the subproblem of \eqref{eq33_1} aims to find a proximal point of $\mathbf{z}_{k-1}$ and usually has a fast closed-form solution depending on $\mathcal{T}$, while the subproblem of \eqref{eq33_2}, from a Bayesian perspective, corresponds to Gaussian denoising on $\mathbf{x}_{k}$ with noise level $\sqrt{\lambda/\mu}$.
Consequently, any Gaussian denoiser can be plugged into the alternating iterations to solve~\eqref{eq2}.
To address this, we rewrite~\eqref{eq33_2} as follows
\begin{equation}\label{eq_priorterm}
  \mathbf{z}_{k} = Denoiser(\mathbf{x}_{k}, \sqrt{\lambda/\mu}).
\end{equation}
One can have two observations from~\eqref{eq_priorterm}. First, the prior $\mathcal{R}(\cdot)$ can be implicitly specified by a denoiser. For this reason, both the prior and the denoiser for plug-and-play IR are usually termed as denoiser prior.
Second, it is interesting to learn a single CNN denoiser to replace~\eqref{eq_priorterm} so as to exploit the advantages of CNN, such as high flexibility of network design, high efficiency on GPUs and powerful modeling capacity with deep networks.

\subsection{General Methodology for Parameter Setting}

From the alternating iterations between \eqref{eq33_1} and \eqref{eq33_2}, it is easy to see that there involves three adjustable parameters, including penalty parameter $\mu$, regularization parameter $\lambda$ and the total number of iterations $K$.

\begin{figure}[!bp]
\vspace{-0.2cm}
\scriptsize{
\begin{center}
\subfigure[$\alpha_k$]
{\includegraphics[width=0.24\textwidth]{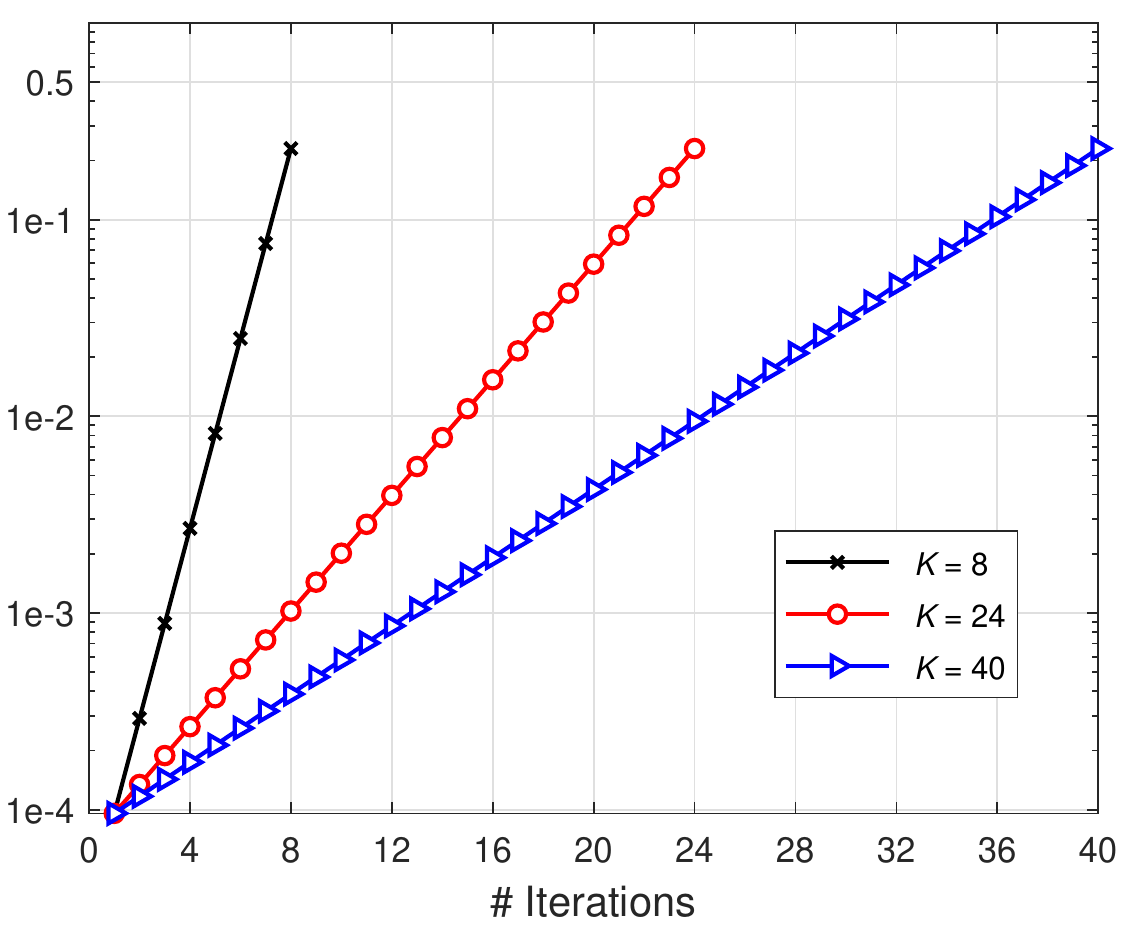}}
\subfigure[$\sigma_k$]
{\includegraphics[width=0.24\textwidth]{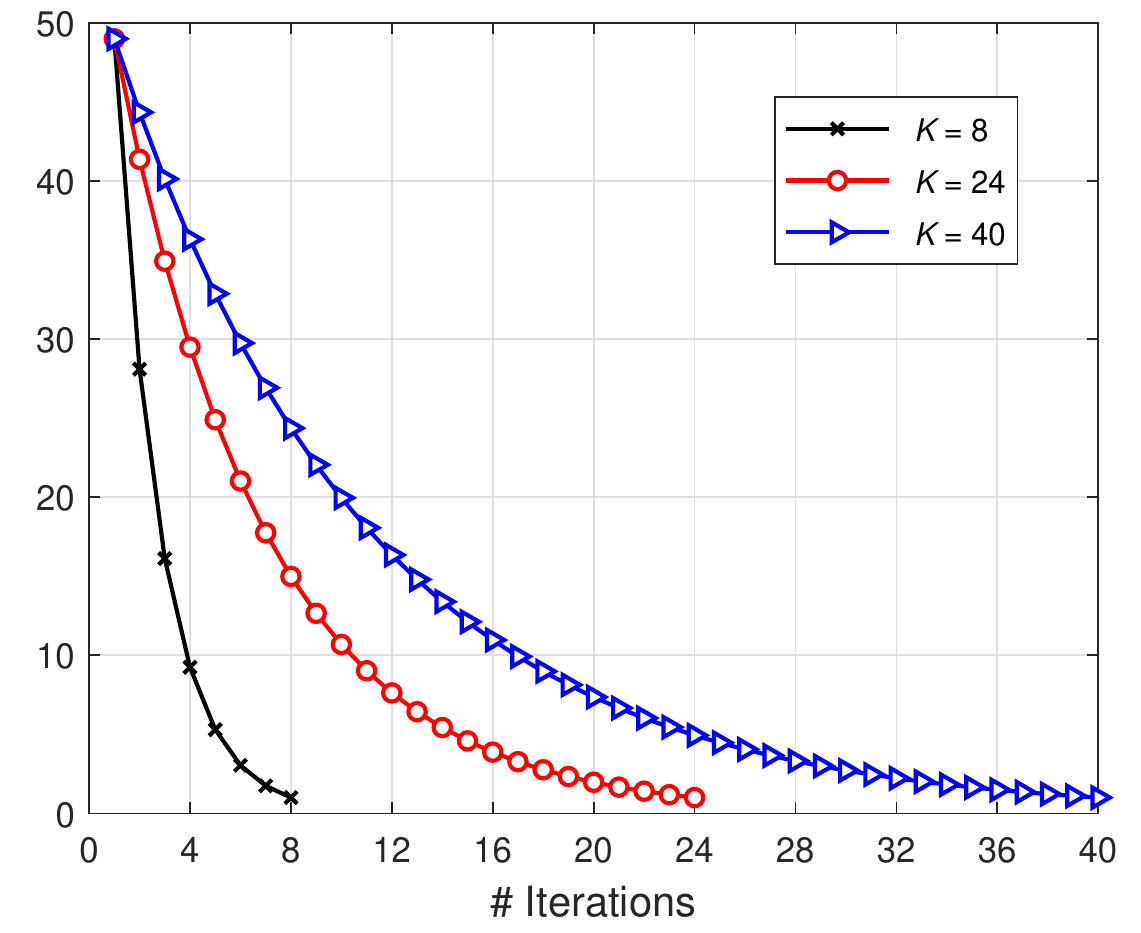}}
\vspace{-0.1cm}
\caption{The values of $\alpha_k$ and $\sigma_k$ at $k$-th iteration with respect to different number of iterations $K$ = 8, 24, and 40.}\label{fig:parameter}
\end{center}}\vspace{-.2cm}
\end{figure}

To guarantee $\mathbf{x}_k$ and $\mathbf{z}_k$ converge to a fixed point, a large $\mu$ is needed, which however requires a large $K$ for convergence. Hence, the common way is to adopt the continuation strategy to gradually increase $\mu$, resulting in a sequence of $\mu_1<\cdots<\mu_k<\cdots<\mu_K$. Nevertheless, a new parameter needs to be introduced to control the step size, making the parameter setting more complicated.
According to~\eqref{eq_priorterm}, we can observe that $\mu$ controls the noise level $\sigma_k(\triangleq\sqrt{\lambda/\mu_k})$ in $k$-th iteration of the denoiser prior. On the other hand, a noise level range of $[0, 50]$ is supposed to be enough for $\sigma_k$. Inspired by such domain knowledge, we can instead set $\sigma_k$ and $\lambda$ to implicitly determine $\mu_k$.
Based on the fact that $\mu_k$ should be monotonically increasing, we uniformly sample $\sigma_k$ from a large noise level $\sigma_{1}$ to a small one $\sigma_{K}$ in log space.
This means that $\mu_k$ can be easily determined via $\mu_k=\lambda/\sigma_k^2$.
Following~\cite{zhang2017learning}, $\sigma_{1}$ is fixed to 49 while $\sigma_K$ is determined by the image noise level $\sigma$. Since $K$ is user-specified and $\sigma_K$ has clear physical meanings, they are practically easy to set.
As for the theoretical convergence of plug-and-play IR, please refer to~\cite{chan2016plug}.

By far, the remaining parameter for setting is $\lambda$. Due to the fact that $\lambda$ comes from the prior term and thus should be fixed, we can choose the optimal $\lambda$ by a grid search on a validation dataset. Empirically, $\lambda$ can yield favorable performance from the range of $[0.19, 0.55]$. In this paper, we fix it to 0.23 unless otherwise specified. It should be noted that since $\lambda$ can be absorbed into $\sigma$ and plays the role of controlling the trade-off between data term and prior term, one can implicitly tune $\lambda$ by multiplying $\sigma$ by a scalar. To have a clear understanding of the parameter setting,
by denoting $\alpha_k \triangleq \mu_k\sigma^2 = {\lambda\sigma^2}/{\sigma_k^2}$ and
assuming $\sigma_K = \sigma=1$, we plot the values of $\alpha_k$ and $\sigma_k$ with respect to different number of iterations $K$ = 8, 24, and 40 in Fig.~\ref{fig:parameter}.

\subsection{Periodical Geometric Self-Ensemble}

Geometric self-ensemble based on flipping and rotation is a commonly-used strategy to boost IR performance~\cite{timofte2016seven}. It first transforms the input via flipping and rotation to generate 8 images, then gets the corresponding restored images after feeding the model with the 8 images, and finally produces the averaged result after the inverse transformation. While a performance gain can be obtained via geometric self-ensemble, it comes at the cost of increased inference time.

Different from the above method, we instead periodically apply the geometric self-ensemble for every successive 8 iterations. In each iteration, there involves one transformation before denoising and the counterpart inverse transformation after denoising. Note that the averaging step is abandoned because the input of the denoiser prior model varies across iterations. We refer to this method as periodical geometric self-ensemble. Its distinct advantage is that the total inference time would not increase. We empirically found that geometric self-ensemble can generally improve the PSNR by 0.02dB$\sim$0.2dB.

Based on the above discussion, we summarized the detailed algorithm of HQS-based plug-and-play IR with deep denoiser prior, namely DPIR, in Algorithm~\ref{alg:dpir}.

\SetCommentSty{textit}
\SetKwComment{tcc}{}{} %default /* */
\SetSideCommentRight
\SetKwInOut{Input}{Input}\SetKwInOut{Output}{Output}
\begin{algorithm}%\small
\Input{Deep denoiser prior model, degraded image $\mathbf{y}$, degradation operation $\mathcal{T}$, image noise level $\sigma$, $\mathrm{\sigma}_k$ of denoiser prior model at $k$-th iteration for a total of $K$ iterations, trade-off parameter $\lambda$.}
 \Output{Restored image $\mathbf{z}_K$.}
\BlankLine
Initialize $\mathbf{z}_{0}$ from $\mathbf{y}$, pre-calculate $\alpha_k \triangleq  {\lambda\sigma^2}/{\sigma_k^2}$.\\
\For{$k = 1, 2, \cdots, K$}
{
$ \mathbf{x}_{k}=\mathop{\arg\min}_{\mathbf{x}} \|\mathbf{y} - \mathcal{T}(\mathbf{x})\|^2 + \alpha_k\|\mathbf{x}-\mathbf{z}_{k-1} \|^2$
\tcc*[l]{\textcolor[rgb]{0.40,0.40,0.40}{// Solving data subproblem}}

$\mathbf{z}_{k} = Denoiser(\mathbf{x}_{k}, \sigma_k)$ \tcc*[l]{\textcolor[rgb]{0.40,0.40,0.40}{// Denoising with deep DRUNet denoiser and periodical geometric self-ensemble}}
%$\mathbf{z}_{k} = G^{-1}(\mathbf{z}_{k}, k\%8)$

}
\caption{Plug-and-play image restoration with deep denoiser prior (DPIR).}\label{alg:dpir}
\end{algorithm}

\section{Experiments}

To validate the effectiveness of the proposed DPIR, we consider three classical IR tasks, including image deblurring, single image super-resolution (SISR), and color image demosaicing.
For each task, we will provide the specific degradation model, fast solution of \eqref{eq33_1} in Algorithm~\ref{alg:dpir}, parameter setting for $K$ and $\sigma_K$, initialization of $\mathbf{z}_0$, and the performance comparison with other state-of-the-art methods.
To further analyze DPIR, we also provide the visual results of $\mathbf{x}_k$ and $\mathbf{z}_k$ at intermediate iterations as well as the convergence curves.
Note that in order to show the advantage of the powerful DRUNet denoiser prior over IRCNN denoiser prior, we refer to DPIR with IRCNN denoiser prior as IRCNN+.

\subsection{Image Deblurring}
The degradation model for deblurring a blurry image with uniform blur (or image deconvolution) is generally expressed as
\begin{equation}\label{eq:deblur}
  \mathbf{y} = \mathbf{x}\otimes \mathbf{k} + \mathbf{n}
\end{equation}
where $\mathbf{x}\otimes \mathbf{k}$ denotes two-dimensional convolution between the latent clean image $\mathbf{x}$ and the blur kernel $\mathbf{k}$.
%Note that we consider the uniform image deblurring which is also termed as image deconvolution.
By assuming the convolution is carried out with circular boundary conditions, the fast solution of \eqref{eq33_1} is given by
\begin{equation}\label{eq:deblur_closedform}
  \mathbf{x}_{k} = \mathcal{F}^{-1}\left(\frac{\overline{\mathcal{F}(\mathbf{k})}\mathcal{F}(\mathbf{y})+\alpha_k\mathcal{F}(\mathbf{z}_{k-1}) }{\overline{\mathcal{F}(\mathbf{k})} \mathcal{F}(\mathbf{k})+\alpha_k}\right)
\end{equation}
where the $\mathcal{F}(\cdot)$ and $\mathcal{F}^{-1}(\cdot)$ denote Fast Fourier Transform (FFT) and inverse FFT, $\overline{\mathcal{F}(\cdot)}$ denotes complex
conjugate of $\mathcal{F}(\cdot)$. It can be noted that the blur kernel $\mathbf{k}$ is only involved in \eqref{eq:deblur_closedform}. In other words, \eqref{eq:deblur_closedform} explicitly handles the distortion of blur.
\begin{figure}[!htbp]
\scriptsize{
\begin{center}
\subfigure[]
{\includegraphics[width=0.077\textwidth]{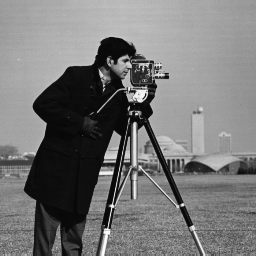}}
\subfigure[]
{\includegraphics[width=0.077\textwidth]{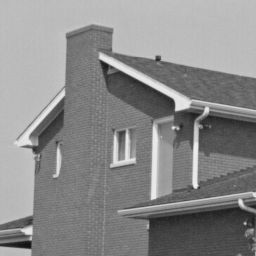}}
\subfigure[]
{\includegraphics[width=0.077\textwidth]{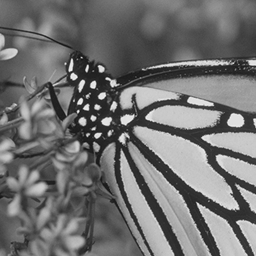}}
\subfigure[]
{\includegraphics[width=0.077\textwidth]{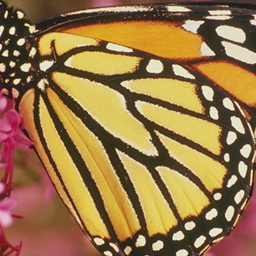}}
\subfigure[]
{\includegraphics[width=0.077\textwidth]{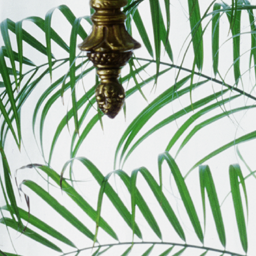}}
\subfigure[]
{\includegraphics[width=0.077\textwidth]{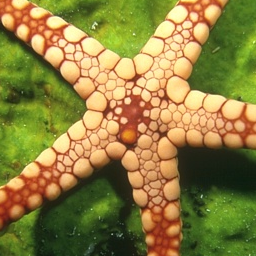}}
\vspace{-0.2cm}
\caption{Six classical testing images. (a) \emph{Cameraman}; (b) \emph{House}; (c) \emph{Monarch}; (d) \emph{Butterfly}; (e) \emph{Leaves}; (f) \emph{Starfish}.}\label{fig:six}
\end{center}}\vspace{-.1cm}
\end{figure}

\begin{table}[!htbp]\footnotesize
\caption{ {PSNR(dB) results of different methods on Set6 for image deblurring.} The best and second best results are highlighted in \textcolor[rgb]{1.00,0.00,0.00}{red} and \textcolor[rgb]{0.00,0.00,1.00}{blue} colors, respectively.}
\vspace{-0.2cm}
\center
\begin{tabular}{p{0.8cm}<{\centering}p{0.5cm}<{\centering}|p{.65cm}<{\centering}p{0.65cm}<{\centering}p{0.74cm}<{\centering}|p{0.7cm}<{\centering}p{0.65cm}<{\centering}p{0.65cm}<{\centering} }%p{0.1\textwidth}
\shline
 \scriptsize Methods& $\sigma$ &\scriptsize \emph{C.man} &\scriptsize \emph{House}   & \scriptsize \emph{Monarch}  & \scriptsize \emph{Butterfly}  &\scriptsize \emph{Leaves}  & \scriptsize \emph{Starfish}   \\ \hline\hline
   \multicolumn{8}{c}{The second kernel of size 17$\times$17 from~\cite{levin2009understanding}}    \\   \hline% \cline{1-7}
   \scriptsize EPLL   &  \multirow{6}{*}{2.55}  & 29.18  & 32.33 & 27.32 & 24.96  &  23.48 & 28.05\\%\cline{2-6}
   \scriptsize FDN  &  & 29.09  & 29.75 & 29.13 & 28.06  & 27.04 & 28.12 \\%\cline{2-6}
    \scriptsize DMPHN  &  & 21.35  & 21.85 & 17.87 & 18.39  & 16.20 & 20.33 \\%\cline{2-6}
 \scriptsize IRCNN  &  & \textcolor[rgb]{0.00,0.00,1.00}{31.69}  & \textcolor[rgb]{0.00,0.00,1.00}{35.04} & \textcolor[rgb]{0.00,0.00,1.00}{32.71} &  \textcolor[rgb]{0.00,0.00,1.00}{33.13} & \textcolor[rgb]{0.00,0.00,1.00}{33.51} & \textcolor[rgb]{0.00,0.00,1.00}{33.15} \\%\cline{2-6}
\scriptsize IRCNN+  &  & 31.23  & 34.01 & 31.85 &  32.55 & 32.66 & 32.34 \\%\cline{2-6}
 \scriptsize  DPIR &  & \textcolor[rgb]{1.00,0.00,0.00}{32.05}  & \textcolor[rgb]{1.00,0.00,0.00}{35.82} & \textcolor[rgb]{1.00,0.00,0.00}{33.38} &  \textcolor[rgb]{1.00,0.00,0.00}{34.26} & \textcolor[rgb]{1.00,0.00,0.00}{35.19} & \textcolor[rgb]{1.00,0.00,0.00}{34.21} \\%\cline{2-6}
 \hline
    \scriptsize EPLL   & \multirow{6}{*}{7.65}    & 24.82  & 28.50  & 23.03 & 20.82  &  20.06 & 24.23\\%\cline{2-6}
    \scriptsize FDN  &  & 26.18  & 28.01 & 25.86 &  24.76 & 23.91 & 25.21 \\%\cline{2-6}
    \scriptsize DMPHN  &  & 20.71  & 22.45 & 19.01 &  17.97 & 15.83 & 19.74 \\%\cline{2-6}
 \scriptsize IRCNN  &  & \textcolor[rgb]{0.00,0.00,1.00}{27.70}  & \textcolor[rgb]{0.00,0.00,1.00}{31.94} & \textcolor[rgb]{0.00,0.00,1.00}{28.23} &  \textcolor[rgb]{0.00,0.00,1.00}{28.73} & \textcolor[rgb]{0.00,0.00,1.00}{28.63} & \textcolor[rgb]{0.00,0.00,1.00}{28.76} \\%\cline{2-6}
 \scriptsize IRCNN+  &  & 27.64  & 31.00 & 27.66 &  28.52 & 28.17 & 28.50 \\%\cline{2-6}
 \scriptsize DPIR  &  & \textcolor[rgb]{1.00,0.00,0.00}{28.17}  & \textcolor[rgb]{1.00,0.00,0.00}{32.79} & \textcolor[rgb]{1.00,0.00,0.00}{28.48} &  \textcolor[rgb]{1.00,0.00,0.00}{29.52} & \textcolor[rgb]{1.00,0.00,0.00}{30.11} & \textcolor[rgb]{1.00,0.00,0.00}{29.83} \\%\cline{2-6}
   \hline
       \multicolumn{8}{c}{The fourth kernel of size 27$\times$27 from~\cite{levin2009understanding}}    \\   \hline% \cline{1-7}
   \scriptsize EPLL   &  \multirow{6}{*}{2.55}  & 27.85  & 28.13 & 22.92 & 20.55  & 19.22 & 24.84\\%\cline{2-6}
   \scriptsize FDN  &  & 28.78  & 29.29 & 28.60 &  27.40 & 26.51 & 27.48 \\%\cline{2-6}
    \scriptsize DMPHN  &  & 16.28  & 17.14 & 12.94 &  11.90 & 9.85 & 17.32 \\%\cline{2-6}
 \scriptsize IRCNN  &  & \textcolor[rgb]{0.00,0.00,1.00}{31.56}  & \textcolor[rgb]{0.00,0.00,1.00}{34.73} & \textcolor[rgb]{0.00,0.00,1.00}{32.42} &  \textcolor[rgb]{0.00,0.00,1.00}{32.74} & 33.22 & \textcolor[rgb]{0.00,0.00,1.00}{32.53} \\%\cline{2-6}
 \scriptsize IRCNN+  &  & 31.29  & 34.17 & 31.82 &  32.48 & \textcolor[rgb]{0.00,0.00,1.00}{33.59} & 32.18 \\%\cline{2-6}
 \scriptsize DPIR &  & \textcolor[rgb]{1.00,0.00,0.00}{31.97}  & \textcolor[rgb]{1.00,0.00,0.00}{35.52} & \textcolor[rgb]{1.00,0.00,0.00}{32.99} &  \textcolor[rgb]{1.00,0.00,0.00}{34.18} & \textcolor[rgb]{1.00,0.00,0.00}{35.12} & \textcolor[rgb]{1.00,0.00,0.00}{33.91} \\%\cline{2-6}
  \hline% \cline{1-7}
   \scriptsize EPLL   &  \multirow{6}{*}{7.65}  & 24.31  & 26.02 & 20.86 & 18.64  &  17.54 & 22.47\\%\cline{2-6}
   \scriptsize FDN  &  & 26.13  & 27.41 & 25.39 &  24.27 & 23.53 & 24.71 \\%\cline{2-6}
  \scriptsize DMPHN  &  & 15.50  & 16.63 & 12.51 &  11.13 & 9.74 & 15.07 \\%\cline{2-6}
 \scriptsize IRCNN  &  & \textcolor[rgb]{0.00,0.00,1.00}{27.58}  & \textcolor[rgb]{0.00,0.00,1.00}{31.55} & \textcolor[rgb]{0.00,0.00,1.00}{27.99} &  \textcolor[rgb]{0.00,0.00,1.00}{28.53} & \textcolor[rgb]{0.00,0.00,1.00}{28.45} & \textcolor[rgb]{0.00,0.00,1.00}{28.42} \\%\cline{2-6}
 \scriptsize IRCNN+  &  & 27.49  & 30.80 & 27.54 &  28.40 & 28.14 & 28.20 \\%\cline{2-6}
 \scriptsize DPIR &  & \textcolor[rgb]{1.00,0.00,0.00}{27.99}  & \textcolor[rgb]{1.00,0.00,0.00}{32.87} & \textcolor[rgb]{1.00,0.00,0.00}{28.27} &  \textcolor[rgb]{1.00,0.00,0.00}{29.45} & \textcolor[rgb]{1.00,0.00,0.00}{30.27} & \textcolor[rgb]{1.00,0.00,0.00}{29.46} \\%\cline{2-6}
  \shline

\end{tabular}
\label{table:deblur}
\end{table}

\begin{figure*}[!tbp]

\begin{center}\hspace{-0.2cm}
\subfigure[Blurry image]
{\includegraphics[width=0.157\textwidth]{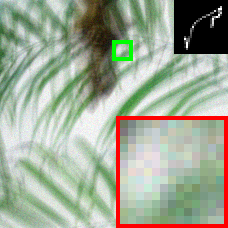}}\hspace{0.12cm}
\subfigure[DMPHN (9.74dB)]
{\includegraphics[width=0.157\textwidth]{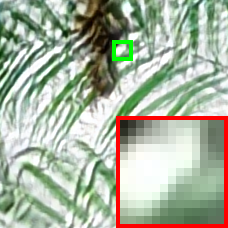}}\hspace{0.12cm}
\subfigure[FDN (23.53dB)]
{\includegraphics[width=0.157\textwidth]{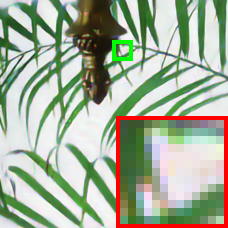}}\hspace{0.12cm}
\subfigure[IRCNN (28.45dB)]
{\includegraphics[width=0.157\textwidth]{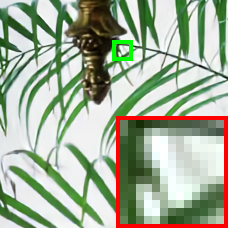}}\hspace{0.12cm}
\subfigure[IRCNN+ (28.14dB)]
{\includegraphics[width=0.157\textwidth]{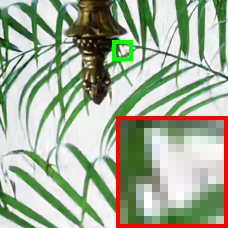}}\hspace{0.12cm}
\subfigure[DPIR (30.27dB)]
{\includegraphics[width=0.157\textwidth]{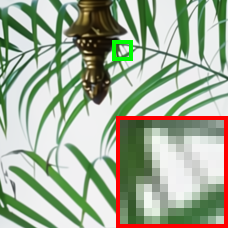}}
\vspace{-0.2cm}
\caption{ {Visual results comparison of different deblurring methods on \emph{Leaves}. The blur kernel is visualized in the upper right corner of the blurry image. The noise level is 7.65(3\%).}}\label{fig:deblur}
\end{center}\vspace{-.2cm}
\end{figure*}

\begin{figure*}[!htbp]
\begin{center}\hspace{-0.2cm}
\subfigure[$\mathbf{x}_1$ (16.34dB)]{
\begin{overpic}[width=0.157\textwidth]{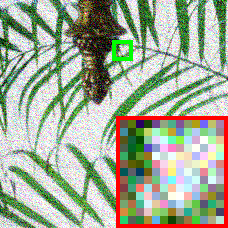}
\end{overpic}}
\subfigure[$\mathbf{z}_1$ (23.75dB)]{
\begin{overpic}[width=0.157\textwidth]{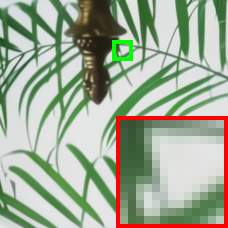}
\end{overpic}}
\subfigure[$\mathbf{x}_4$ (22.33dB)]{
\begin{overpic}[width=0.157\textwidth]{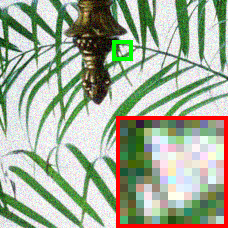}
\end{overpic}}
\subfigure[$\mathbf{z}_4$ (29.37dB)]{
\begin{overpic}[width=0.157\textwidth]{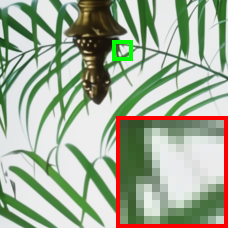}
\end{overpic}}
\subfigure[$\mathbf{x}_8$ (29.34dB)]{
\begin{overpic}[width=0.157\textwidth]{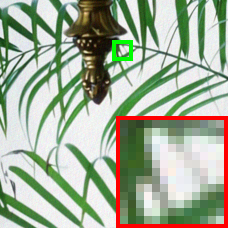}
\end{overpic}}
\subfigure[Convergence curves]{
\begin{overpic}[width=0.157\textwidth]{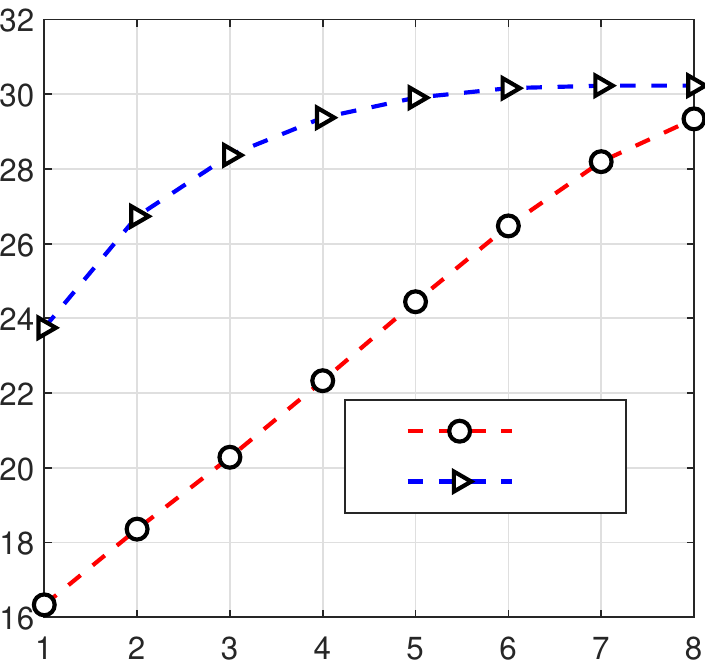}
\put(76,24.8){\color{black}{\tiny $\mathbf{z}_k$}}
\put(76,32){\color{black}{\tiny $\mathbf{x}_k$}}
\end{overpic}}
\vspace{-0.4cm}
\caption{(a)-(e) Visual results and PSNR results of $\mathbf{x}_k$ and $\mathbf{z}_k$ at different iterations; (f) Convergence curves of PSNR  (y-axis) for $\mathbf{x}_k$ and $\mathbf{z}_k$  with respect to number of iterations (x-axis).}
\label{fig:deblur_iter}
\end{center}\vspace{-.2cm}
\end{figure*}

\subsubsection{Quantitative and Qualitative Comparison}

For the sake of making a quantitative analysis on the proposed DPIR, we consider six classical testing images as shown in Fig.~\ref{fig:six} and two of the eight real blur kernels from~\cite{levin2009understanding}. Specifically, the testing images which we refer to as Set6 consist of 3 grayscale images and 3 color images. Among them, \emph{House} and \emph{Leaves} are full of repetitive structures and thus can be used to evaluate non-local self-similarity prior.
For the two blur kernels, they are of size 17$\times$17 and 27$\times$27, respectively.
As shown in Table~\ref{table:deblur}, we also consider Gaussian noise with different noise levels 2.55(1\%) and 7.65(3\%). Following the common setting, we synthesize the blurry images by first applying a blur kernel and then adding AWGN with noise level $\sigma$. For the parameters $K$ and $\sigma_K$, they are set to 8 and $\sigma$, respectively. For $\mathbf{z}_0$, it is initialized as $\mathbf{y}$.

 {To evaluate the effectiveness of the proposed DPIR, we choose four representative methods for comparison, including model-based method EPLL~\cite{zoran2011learning}, learning-based non-blind method FDN~\cite{kruse2017learning}, learning-based blind method DMPHN~\cite{zhang2019deep} and plug-and-play method IRCNN and IRCNN+. Table~\ref{table:deblur} summarizes the PSNR results on Set6.
As one can see, DMPHN obtains the lowest PSNR values possibly due to the lacking of the FFT module. In contrast, the proposed DPIR outperforms EPLL and FDN by a large margin. Although DPIR has 8 iterations rather than 30 iterations of IRCNN, it has a PSNR gain of 0.2dB$\sim$2dB over IRCNN. On the other hand, with the same number of iterations, DPIR significantly outperforms IRCNN+, which indicates that the denoiser plays a vital role in plug-and-play IR. In addition, one can see that the PSNR gain of DPIR over IRCNN and IRCNN+ on \emph{House} and \emph{Leaves} is larger than those on other images. A possible reason is that the DRUNet denoiser learns more nonlocal self-similarity prior than the shallow denoisers of IRCNN.}

 {The visual comparison of different methods on \emph{Leaves} with the fourth kernel and noise level 7.65 is shown in Fig.~\ref{fig:deblur}. We can see that DMPHN can remove the noise but fail to recover the image sharpness, while FDN tends to smooth out fine details and
generate color artifacts.} Although IRCNN and IRCNN+ avoid the color artifacts, it fails to recover the fine details. In contrast, the proposed DPIR can recover image sharpness and naturalness.

\subsubsection{Intermediate Results and Convergence}

Figs.~\ref{fig:deblur_iter}(a)-(e) provide the visual results of $\mathbf{x}_k$ and $\mathbf{z}_k$ at different iterations on the testing image from Fig.~\ref{fig:deblur}, while Fig.~\ref{fig:deblur_iter}(f) shows the PSNR convergence curves for $\mathbf{x}_k$ and $\mathbf{z}_k$.
We can have the following observations. First, while \eqref{eq33_1} can handle the distortion of blur, it also aggravates the strength of noise compared to its input $\mathbf{z}_{k-1}$. Second, the deep denoiser prior plays the role of removing noise, leading to a noise-free $\mathbf{z}_{k}$. Third, compared with $\mathbf{x}_1$ and $\mathbf{x}_2$, $\mathbf{x}_8$ contains more fine details, which means \eqref{eq33_1} can iteratively recover the details. Fourth, according to Fig.~\ref{fig:deblur_iter}(f), $\mathbf{x}_k$ and $\mathbf{z}_k$ enjoy a fast convergence to the fixed point.

\subsubsection{Analysis of the Parameter Setting}\label{sec:parametersetting}

While we fixed the total number of iterations $K$ to be 8 and the noise level in the first iteration $\sigma_1$ to be 49, it is interesting to investigate the performance with other settings. Table~\ref{table:parametersetting} reports the PSNR results with different combinations of $K$ and $\sigma_1$ on the testing image from Fig.~\ref{fig:deblur}. One can see that larger $\sigma_1$, such as 39 and 49, could result in better PSNR results. On the other hand, if $\sigma_1$ is small, a large $K$ needs to be specified for a good performance, which however would increase the computational burden. As a result, $K$ and $\sigma_1$ play an important role for the trade-off between efficiency and effectiveness.

\begin{table}[!htbp]\footnotesize
\caption{PSNR results with different combinations of $K$ and $\sigma_1$ on the testing image from Fig.~\ref{fig:deblur}.}
%\vspace{-0.3cm}
\center
\begin{tabular}{p{0.8cm}<{\centering}|p{1.1cm}<{\centering}p{1.1cm}<{\centering}p{1.1cm}<{\centering}p{1.1cm}<{\centering}p{1.1cm}<{\centering}}
\shline
\multirow{2}{*}{$K$}  &  \multirow{2}{*}{$\sigma_1 = 9$}  &\multirow{2}{*}{$\sigma_1 = 19$}  & \multirow{2}{*}{$\sigma_1 = 29$} & \multirow{2}{*}{$\sigma_1 = 39$} & \multirow{2}{*}{$\sigma_1 = 49$}  \\
  &  &   &  &  &    \\\hline\hline
  4 & 20.04 & 23.27  & 25.70 & 27.65 & 28.96   \\
  8 & 22.50 & 25.96  & 28.40 & 29.89 & 30.27   \\
 24 & 26.58 & 29.64  & 30.06 & 30.13 & 30.16  \\
 40 & 28.60 & 29.82  & 29.92 & 29.98 & 30.01  \\
  \shline
\end{tabular}
\label{table:parametersetting}
\end{table}

\subsubsection{ {Results of DPIR with Blind DRUNet Denoiser}}\label{sec:blinddenoiser}
\begin{figure}[!htbp]
\begin{center}\hspace{-0.2cm}
\subfigure[$\mathbf{x}_4$ (21.66dB)]{
\begin{overpic}[width=0.156\textwidth]{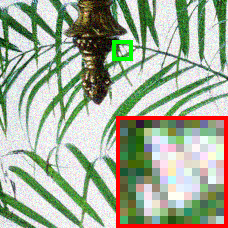}
\end{overpic}}
\subfigure[$\mathbf{z}_4$ (27.64dB)]{
\begin{overpic}[width=0.156\textwidth]{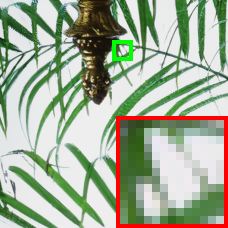}
\end{overpic}}
\subfigure[$\mathbf{z}_8$ (25.59dB)]{
\begin{overpic}[width=0.156\textwidth]{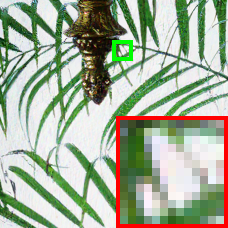}
\end{overpic}}
\vspace{-0.4cm}
\caption{ {Visual results and PSNR results of $\mathbf{x}_k$ and $\mathbf{z}_k$ at different iterations of the proposed DPIR with blind DRUNet denoiser. The testing image is same as in Figs.~\ref{fig:deblur} and~\ref{fig:deblur_iter}.}}
\label{fig:deblur_iter_blind}
\end{center}\vspace{-.2cm}
\end{figure}

 {It is interesting to demonstrate the performance of DPIR with blind DRUNet denoiser. For this purpose, we have trained a blind DRUNet denoiser by removing the noise level map. With the same parameter setting, we provide the visual results and PSNR results of $\mathbf{x}_k$ and $\mathbf{z}_k$ at different iterations of DPIR with the blind DRUNet denoiser on the noisy and blurry \emph{Leaves} in Fig.~\ref{fig:deblur_iter_blind}. Note that the testing image is same as in Figs.~\ref{fig:deblur} and~\ref{fig:deblur_iter}. By comparison, we can see that DPIR with blind denoiser gives rise to much lower PSNR values than DPIR with non-blind denoiser. In particular, some structural artifacts can be observed with a closer look at the final result $\mathbf{z}_8$. 
As a result, the non-blind denoiser is more suitable than blind denoiser for plug-and-play IR.}

\subsection{Single Image Super-Resolution (SISR)}

While existing SISR methods are mainly designed for bicubic degradation model with the following formulation
\begin{equation}\label{eq:sisr_bicubic_degradation}
  \mathbf{y} = \mathbf{x}\downarrow^{bicubic}_s,
\end{equation}
where $\downarrow^{bicubic}_s$ denotes bicubic downsamling with downscaling factor $s$, it has been revealed that these methods would deteriorate seriously if the real degradation model deviates from the assumed one~\cite{efrat2013accurate,zhang2015revisiting}. To remedy this, an alternative way is to adopt a classical but practical degradation model which assumes the low-resolution (LR) image is a blurred, decimated, and noisy version of high-resolution (HR) image. The mathematical formulation of such degradation model is given by
\begin{equation}\label{eq:sisr_degradation}
  \mathbf{y} = (\mathbf{x}\otimes \mathbf{k})\downarrow_s + ~\mathbf{n},
\end{equation}
where $\downarrow_s$ denotes the standard $s$-fold downsampler, i.e., selecting the upper-left pixel for each distinct $s$$\times$$s$ patch.

In this paper, we consider the above-mentioned two degradation models for SISR. As for the solution of \eqref{eq33_1}, the following iterative back-projection (IBP) solution~\cite{irani1993motion,egiazarian2015single} can be adopted for bicubic degradation,
\begin{equation}\label{eq:sisr_backprojection}
  \mathbf{x}_{k} = \mathbf{z}_{k-1} - \gamma(\mathbf{y} - \mathbf{z}_{k-1}\downarrow^{bicubic}_s)\uparrow^{bicubic}_s,
\end{equation}
where $\uparrow^{bicubic}_s$ denotes bicubic interpolation with upscaling factor $s$, $\gamma$ is the step size. Note that we only show one iteration for simplicity.
As an extension, \eqref{eq:sisr_backprojection} can be further modified as follows to handle the classical degradation model
 \begin{equation}\label{eq:sisr_bp}
  \mathbf{x}_{k} = \mathbf{z}_{k-1} - \gamma\Big((\mathbf{y} - (\mathbf{z}_{k-1}\otimes \mathbf{k})\downarrow_s)\uparrow_s\Big)\otimes \mathbf{k},
\end{equation}
where $\uparrow_s$ denotes upsampling the spatial size by filling the new entries with zeros.
Especially noteworthy is that there exists a fast close-form solution to replace the above iterative scheme. According to~\cite{zhao2016fast},
by assuming the convolution is carried out with circular boundary conditions as in deblurring, the closed-form solution is given by
\begin{equation}\label{eq:sisr}
  \mathbf{x}_{k} = \mathcal{F}^{-1}\left(\frac{1}{\alpha_k}\Big(\mathbf{d} - \overline{\mathcal{F}(\mathbf{k})} \odot_s\frac{(\mathcal{F}(\mathbf{k})\mathbf{d})\Downarrow_s }{(\overline{\mathcal{F}(\mathbf{k})}\mathcal{F}(\mathbf{k}))\Downarrow_s +\alpha_k}\Big)\right),
\end{equation}
where $\mathbf{d} = \overline{\mathcal{F}(\mathbf{k})}\mathcal{F}(\mathbf{y}\uparrow_s) + \alpha_{k}\mathcal{F}(\mathbf{z}_{k-1})$
and where $\odot_s$ denotes distinct block processing operator with element-wise multiplication, i.e., applying element-wise multiplication to the $s\times s$ distinct blocks of $\overline{\mathcal{F}(\mathbf{k})}$, $\Downarrow_s$ denotes distinct block downsampler, i.e., averaging the $s\times s$ distinct blocks~\cite{zhang2020deep}.
It is easy to verify that \eqref{eq:deblur} is a special case of \eqref{eq:sisr} with $s=1$. It is worth noting that \eqref{eq:sisr_degradation} can also be used to solve bicubic degradation by setting the blur kernel to the approximated bicubic kernel~\cite{zhang2020deep}.
In general, the closed-form solution~\eqref{eq:sisr} should be advantageous over iterative solutions~\eqref{eq:sisr_bp}. The reason is that the former is an exact solution which contains one parameter (i.e., $\alpha_k$) whereas the latter is an inexact solution which involves two parameters  (i.e., number of inner iterations per outer iteration and step size).

For the overall parameter setting, $K$ and $\sigma_K$ are set to 24 and $\max(\sigma, s)$, respectively. For the parameters in~\eqref{eq:sisr_backprojection} and~\eqref{eq:sisr_bp}, $\gamma$ is fixed to 1.75, the the number of inner iterations required per outer iteration is set to 5. For the initialization of $z_0$, the bicubic interpolation of the LR image is utilized. In particular, since the classical degradation model selects the upper-left pixel for each distinct $s\times s$ patch, a shift problem should be properly addressed.  {To tackle with this, we adjust $z_0$ by using 2D linear grid interpolation.}

\begin{figure}[!htbp]
\scriptsize{
\begin{center}
\subfigure[]
{\includegraphics[width=0.057\textwidth]{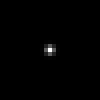}}
\subfigure[]
{\includegraphics[width=0.057\textwidth]{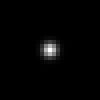}}
\subfigure[]
{\includegraphics[width=0.057\textwidth]{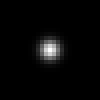}}
\subfigure[]
{\includegraphics[width=0.057\textwidth]{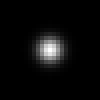}}
\subfigure[]
{\includegraphics[width=0.057\textwidth]{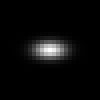}}
\subfigure[]
{\includegraphics[width=0.057\textwidth]{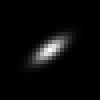}}
\subfigure[]
{\includegraphics[width=0.057\textwidth]{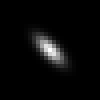}}
\subfigure[]
{\includegraphics[width=0.057\textwidth]{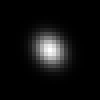}}
\vspace{-0.2cm}
\caption{The eight testing Gaussian kernels for SISR. (a)-(d) are isotropic Gaussian kernels; (e)-(f) are anisotropic Gaussian kernels.}\label{fig:kernels}
\end{center}}%\vspace{-.2cm}
\end{figure}

\subsubsection{Quantitative and Qualitative Comparison}

\begin{figure*}[!htbp]
\begin{center}%\hspace{-0.25cm}
\subfigure[Bicubic (24.82dB)]{
\begin{overpic}[width=0.157\textwidth]{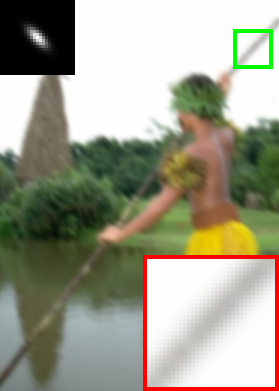}
\end{overpic}}
\subfigure[RCAN (24.61dB)]{
\begin{overpic}[width=0.157\textwidth]{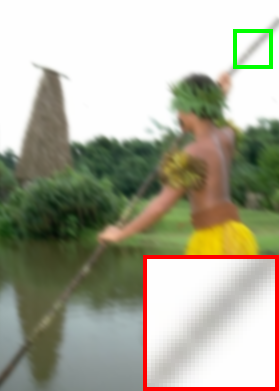}
\end{overpic}}
\subfigure[MZSR (27.34dB)]{
\begin{overpic}[width=0.157\textwidth]{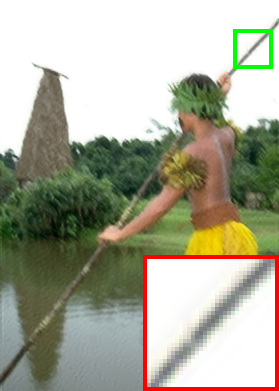}
\end{overpic}}
\subfigure[IRCNN (26.89dB)]{
\begin{overpic}[width=0.157\textwidth]{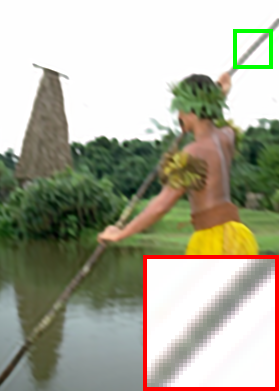}
\end{overpic}}
\subfigure[IRCNN+ (28.65dB)]{
\begin{overpic}[width=0.157\textwidth]{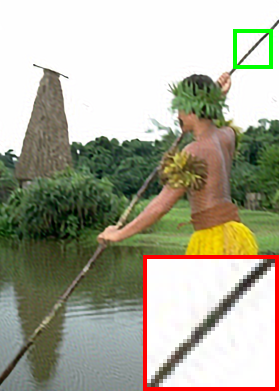}
\end{overpic}}
\subfigure[DPIR (29.12dB)]{
\begin{overpic}[width=0.157\textwidth]{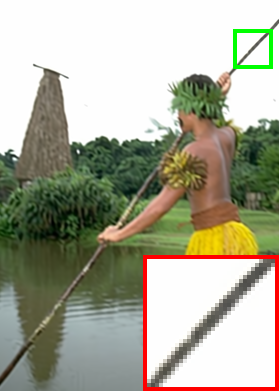}
\end{overpic}}\vspace{-.2cm}
\caption{Visual results comparison of different SISR methods on an image corrupted by classical degradation model. The kernel is shown on the upper-left corner of the bicubicly interpolated LR image. The scale factor is 2.}\label{fig:sr}
\end{center}\vspace{-.2cm}
\end{figure*}

\begin{figure*}[!htbp]
\centering\hspace{-0.25cm}
\subfigure[$\mathbf{x}_1$ (24.95dB)]{
\begin{overpic}[width=0.157\textwidth]{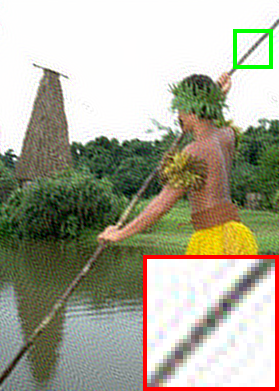}
\end{overpic}}
\subfigure[$\mathbf{z}_1$ (27.24dB)]{
\begin{overpic}[width=0.157\textwidth]{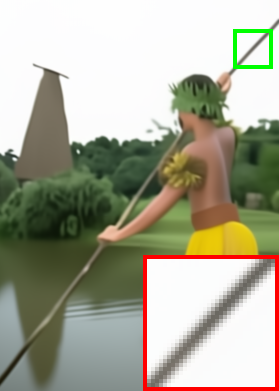}
\end{overpic}}
\subfigure[$\mathbf{x}_6$ (27.59dB)]{
\begin{overpic}[width=0.157\textwidth]{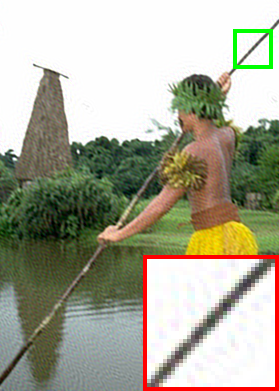}
\end{overpic}}
\subfigure[$\mathbf{z}_6$ (28.57dB)]{
\begin{overpic}[width=0.157\textwidth]{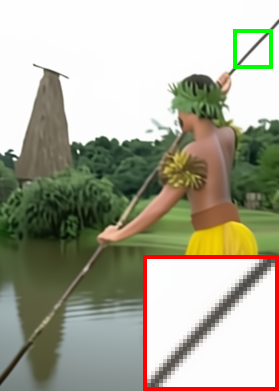}
\end{overpic}}
\subfigure[$\mathbf{x}_{24}$ (29.12dB)]{
\begin{overpic}[width=0.157\textwidth]{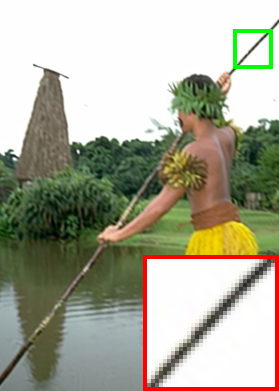}
\end{overpic}}
\subfigure[Convergence curves]{
\begin{overpic}[width=0.157\textwidth]{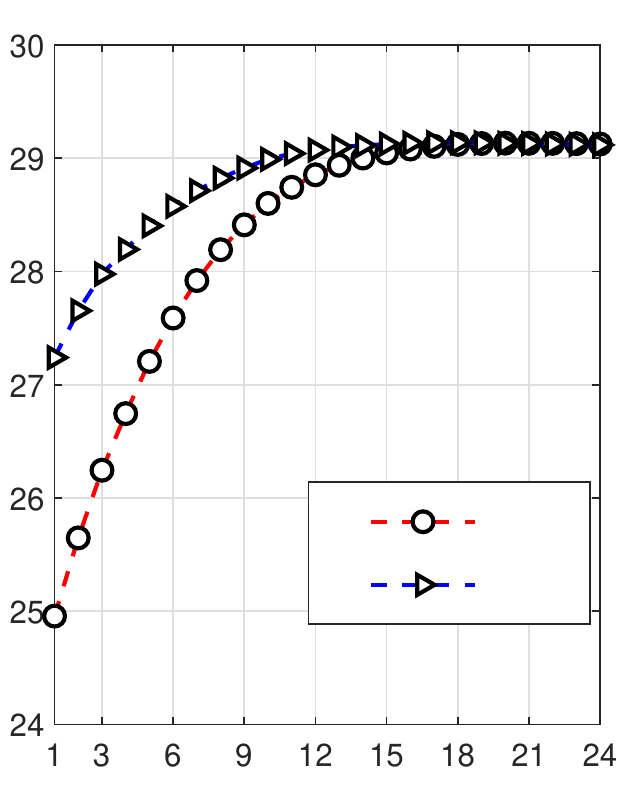}
\put(63,24.6){\color{black}{\tiny $\mathbf{z}_k$}}
\put(63,32.6){\color{black}{\tiny $\mathbf{x}_k$}}
\end{overpic}}
\vspace{-0.2cm}
\caption{(a)-(e) Visual results and PSNR results of $\mathbf{x}_k$ and $\mathbf{z}_k$ at different iteration; (f) Convergence curves of PSNR results (y-axis) for $\mathbf{x}_k$ and $\mathbf{z}_k$  with respect to number of iterations (x-axis).}
\label{fig:sr_iter}\vspace{-.2cm}
\end{figure*}

\begin{table*}[!htbp]\footnotesize
\caption{{Average PSNR(dB) results of different methods for single image super-resolution on CBSD68 dataset.} The best and second best results are
highlighted in \textcolor[rgb]{1.00,0.00,0.00}{red} and \textcolor[rgb]{0.00,0.00,1.00}{blue} colors, respectively.} \vspace{-0.2cm}
\center
\begin{tabular}{p{0.9cm}<{\centering}|p{0.42cm}<{\centering}p{0.42cm}<{\centering}p{0.42cm}<{\centering}p{0.45cm}<{\centering}p{0.45cm}<{\centering}p{0.55cm}<{\centering}p{0.52cm}<{\centering}|p{0.42cm}<{\centering}p{0.42cm}<{\centering}p{0.42cm}<{\centering}p{0.45cm}<{\centering}p{0.55cm}<{\centering}p{0.52cm}<{\centering}|p{0.42cm}<{\centering}p{0.42cm}<{\centering}p{0.42cm}<{\centering}p{0.45cm}<{\centering}p{0.55cm}<{\centering}p{0.5cm}<{\centering}}
  \shline
\multirow{2}{*}{\scriptsize Methods} & \multirow{2}{*}{\scriptsize Bicubic} & \multirow{2}{*}{\scriptsize RCAN} &\multirow{2}{*}{\scriptsize SRFBN} &\multirow{2}{*}{\scriptsize MZSR} & \multirow{2}{*}{\scriptsize IRCNN} & \multirow{2}{*}{\scriptsize IRCNN+} & \multirow{2}{*}{\scriptsize DPIR} & \multirow{2}{*}{\scriptsize Bicubic} & \multirow{2}{*}{\scriptsize RCAN} &\multirow{2}{*}{\scriptsize SRFBN} & \multirow{2}{*}{\scriptsize IRCNN}& \multirow{2}{*}{\scriptsize IRCNN+} & \multirow{2}{*}{\scriptsize DPIR}  &  \multirow{2}{*}{\scriptsize Bicubic} & \multirow{2}{*}{\scriptsize RCAN} &\multirow{2}{*}{\scriptsize SRFBN} & \multirow{2}{*}{\scriptsize IRCNN} &\multirow{2}{*}{\scriptsize IRCNN+} & \multirow{2}{*}{\scriptsize DPIR} \\
 & &  &     &  &  &  &  & &  &  &  &   &  &  &  & &  &  & \\ \hline \hline

\scriptsize Kernel & \multicolumn{7}{c|}{$s$ = 2, $\sigma=0$} & \multicolumn{6}{c|}{$s$ = 3, $\sigma=0$}  & \multicolumn{6}{c}{$s$ = 3, $\sigma=7.65 (3\%)$} \\ \hline

(a)   & 27.60 & 29.50&29.39 & 28.89 & \textcolor[rgb]{0.00,0.00,1.00}{29.92} & \textcolor[rgb]{1.00,0.00,0.00}{30.00}  & 29.78 & 25.83 & 25.02 & 24.98 & 26.43 & \textcolor[rgb]{1.00,0.00,0.00}{26.56} & \textcolor[rgb]{0.00,0.00,1.00}{26.50} & 24.65 &22.77 & 22.58 & 25.45 & \textcolor[rgb]{0.00,0.00,1.00}{25.58} & \textcolor[rgb]{1.00,0.00,0.00}{26.01}\\
(b)   & 26.14 & 26.77 &26.75  & 29.45 & 29.49 & \textcolor[rgb]{1.00,0.00,0.00}{30.28} & \textcolor[rgb]{0.00,0.00,1.00}{30.16} & 25.57 & 27.37 & 27.31 & 26.88 & \textcolor[rgb]{1.00,0.00,0.00}{27.12} & \textcolor[rgb]{0.00,0.00,1.00}{27.08} & 24.45 &24.01 & 24.04 & 25.28 & \textcolor[rgb]{0.00,0.00,1.00}{25.30} & \textcolor[rgb]{1.00,0.00,0.00}{26.10}\\
(c)   & 25.12 & 25.32 &25.31  & 28.49 & 27.75 & \textcolor[rgb]{0.00,0.00,1.00}{29.23}  & \textcolor[rgb]{1.00,0.00,0.00}{29.72} & 24.92 & 25.87 & 25.84 & 26.56 & \textcolor[rgb]{1.00,0.00,0.00}{27.23} & \textcolor[rgb]{0.00,0.00,1.00}{27.21} & 23.94 &23.42 & 23.45 & 24.61 & \textcolor[rgb]{0.00,0.00,1.00}{24.92} & \textcolor[rgb]{1.00,0.00,0.00}{25.71}\\
(d)   & 24.31 & 24.37 &24.35  & 25.26 & 26.44 & \textcolor[rgb]{0.00,0.00,1.00}{27.82}  & \textcolor[rgb]{1.00,0.00,0.00}{28.71} & 24.27 & 24.69 & 24.70 & 25.78 & \textcolor[rgb]{0.00,0.00,1.00}{27.08} & \textcolor[rgb]{1.00,0.00,0.00}{27.18} & 23.41 &22.76 & 22.79 & 23.97 & \textcolor[rgb]{0.00,0.00,1.00}{24.63} & \textcolor[rgb]{1.00,0.00,0.00}{25.17}\\
(e)   & 24.29 & 24.38 &24.37  & 25.48 & 26.41 & \textcolor[rgb]{0.00,0.00,1.00}{27.76}  & \textcolor[rgb]{1.00,0.00,0.00}{28.40} & 24.20 & 24.65 & 24.64 & 25.55 & \textcolor[rgb]{0.00,0.00,1.00}{26.78} & \textcolor[rgb]{1.00,0.00,0.00}{27.05} & 23.35 &22.71 & 22.73 & 23.96 & \textcolor[rgb]{0.00,0.00,1.00}{24.58} & \textcolor[rgb]{1.00,0.00,0.00}{25.09}\\
(f)   & 24.02 & 24.10 &24.09  & 25.46 & 26.05 & \textcolor[rgb]{0.00,0.00,1.00}{27.72}  & \textcolor[rgb]{1.00,0.00,0.00}{28.50} & 23.98 & 24.46 & 24.43 & 25.44 & \textcolor[rgb]{0.00,0.00,1.00}{26.87} & \textcolor[rgb]{1.00,0.00,0.00}{27.04} & 23.16 &22.54 & 22.55 & 23.75 & \textcolor[rgb]{0.00,0.00,1.00}{24.51} & \textcolor[rgb]{1.00,0.00,0.00}{25.01}\\
(g)   & 24.16 & 24.24 & 24.22 & 25.93 & 26.28 & \textcolor[rgb]{0.00,0.00,1.00}{27.86}  & \textcolor[rgb]{1.00,0.00,0.00}{28.66} & 24.10 & 24.63 & 24.61 & 25.64 & \textcolor[rgb]{0.00,0.00,1.00}{27.00} & \textcolor[rgb]{1.00,0.00,0.00}{27.11} & 23.27 &22.64 & 22.68 & 23.87 & \textcolor[rgb]{0.00,0.00,1.00}{24.59} & \textcolor[rgb]{1.00,0.00,0.00}{25.12}\\
(h)   & 23.61 & 23.61 &23.59  & 22.27 & 25.45 & \textcolor[rgb]{0.00,0.00,1.00}{26.88} & \textcolor[rgb]{1.00,0.00,0.00}{27.57} & 23.63 & 23.82 & 23.82 & 24.92 & \textcolor[rgb]{0.00,0.00,1.00}{26.55} & \textcolor[rgb]{1.00,0.00,0.00}{26.94} & 22.88 &22.18 & 22.20 & 23.41 & \textcolor[rgb]{0.00,0.00,1.00}{24.27} & \textcolor[rgb]{1.00,0.00,0.00}{24.60}\\\hline
\scriptsize Bicubic   & 26.37 & \textcolor[rgb]{1.00,0.00,0.00}{31.18} & \textcolor[rgb]{0.00,0.00,1.00}{31.07} & 29.47 & 30.31 &  30.34 & 30.12 & 25.97 & \textcolor[rgb]{1.00,0.00,0.00}{28.08} & \textcolor[rgb]{0.00,0.00,1.00}{28.00} & 27.19 & 27.24 & 27.23 & 24.76 & 24.21 & 24.24 & 24.36 & \textcolor[rgb]{0.00,0.00,1.00}{25.61} & \textcolor[rgb]{1.00,0.00,0.00}{26.35}\\
\shline
\end{tabular}
\label{table:SR}%\vspace{-.2cm}
\end{table*}

To evaluate the flexibility of DPIR, we consider bicubic degradation model, and classical degradation model with 8 diverse Gaussian blur kernels as shown in Fig.~\ref{fig:kernels}.
Following~\cite{zhang2020deep}, the 8 kernels consist of 4 isotropic kernels with different standard deviations (i.e., 0.7, 1.2, 1.6 and 2.0) and 4 anisotropic kernels.
We do not consider motion blur kernels since it has been pointed out that Gaussian kernels are enough for SISR task. To further analyze the performance, three different combinations of scale factor and noise level, including ($s$ = $2$, $\sigma$ = $0$), ($s$ = $3$, $\sigma$ = $0$) and ($s$ = $3$, $\sigma$ = $7.65$), are considered.

For the compared methods, we consider the bicubic interpolation method, RCAN~\cite{zhang2018image}, SRFBN~\cite{li2019feedback}, MZSR~\cite{soh2020meta}, IRCNN and IRCNN+. Specifically, RCAN is the state-of-the-art bicubic degradation based deep model consisting of about 400 layers. SRFBN is a recurrent neural network with feed-back mechanism.
Note that we do not retrain the RCAN and SRFBN models to handle the testing degradation cases as they lack flexibility. Moreover, it is unfair because our DPIR can handle a much wider range of degradations.
MZSR is a zero-shot method based on meta-transfer learning which learns an initial network and then fine-tunes the model on a pair of given LR image and its re-degraded LR image with a few gradient updates.
Similar to IRCNN and DPIR, MZSR is a non-blind method that assumes the blur kernel is known beforehand.
Since MZSR needs to downsample the LR image for fine-tuning, the scale factor should be not too large in order to capture enough information. As a result, MZSR is mainly designed for scale factor 2.

Table~\ref{table:SR} reports the average PSNR(dB) results of different methods for bicubic degradation and classical degradation on color BSD68 dataset.
From Table~\ref{table:SR}, we can have several observations.   {First, as expected, RCAN and SRFBN achieve promising results on bicubic degradation with $\sigma$ = 0 but lose effectiveness when the true degradation deviates from the assumed one. We note that SAN~\cite{dai2019second} has a similar performance to RCAN and SRFBN since they are trained for bicubic degradation.} Second, with the accurate classical degradation model, MZSR outperforms RCAN on most of the blur kernels. Third, IRCNN has a clear PSNR gain over MZSR on smoothed blur kernel. The reason is that MZSR relies heavily on the internal learning of LR image. Fourth, IRCNN performs better on bicubic kernel and the first isotropic Gaussian kernel with noise level $\sigma=0$ than others. This indicates that the IBP solution has very limited generalizability. On the other hand, IRCNN+ has a much higher PSNR than IRCNN, which demonstrates the advantage of closed-form solution over the IBP solution. Last, DPIR can further improves over IRCNN+ by using a more powerful denoiser.
\begin{table*}[!htbp]\footnotesize
\caption{Demosaicing results of different methods on Kodak and McMaster datasets. The best and second best results are
highlighted in \textcolor[rgb]{1.00,0.00,0.00}{red} and \textcolor[rgb]{0.00,0.00,1.00}{blue} colors, respectively.} %\vspace{-0.2cm}
\center
\resizebox{\linewidth}{!}
{
\begin{tabular}{p{1.2cm}<{\centering}|p{1.cm}<{\centering}p{1.cm}<{\centering}p{1cm}<{\centering}p{1.0cm}<{\centering}p{1.cm}<{\centering}p{1.0cm}<{\centering}|p{1.0cm}<{\centering}p{1.0cm}<{\centering}p{1.0cm}<{\centering}p{1.0cm}<{\centering}p{1.0cm}<{\centering}p{1.0cm}<{\centering}}
  \shline
 \multirow{2}{*}{Datasets} &   \multirow{2}{*}{Matlab}& \multirow{2}{*}{DDR} & \multirow{2}{*}{DeepJoint}  & \multirow{2}{*}{MMNet}& \multirow{2}{*}{RLDD}  & \multirow{2}{*}{RNAN}& \multirow{2}{*}{LSSC}& \multirow{2}{*}{IRI}& \multirow{2}{*}{FlexISP} & \multirow{2}{*}{IRCNN} &\multirow{2}{*}{IRCNN+} & \multirow{2}{*}{DPIR}\\%\cline{4-10}

 &  &  &    &   &   &  &  &  &   & & \\ \hline\hline

 Kodak     & 35.78 & 41.11  & 42.00 & 40.19 & 42.49 & \textcolor[rgb]{1.00,0.00,0.00}{43.16} &  41.43 & 39.23 & 38.52  & 40.29 & 40.80 &   \textcolor[rgb]{0.00,0.00,1.00}{42.68} \\
 McMaster  & 34.43 & 37.12  & 39.14 & 37.09 & 39.25 & \textcolor[rgb]{1.00,0.00,0.00}{39.70} &  36.15 & 36.90 & 36.87  & 37.45 & 37.79 & \textcolor[rgb]{0.00,0.00,1.00}{39.39}\\\shline
\end{tabular}}
\label{table_demosaicing}%\vspace{-0.1cm}
\end{table*}

Fig.~\ref{fig:sr} shows the visual comparison of different SISR methods on an image corrupted by classical degradation model. It can be observed that MZSR and IRCNN produce better visual results than bicubic interpolation method. With an inaccurate data term solution, IRCNN fails to recover sharp edges. In comparison,
by using a closed-form data term solution, IRCNN+ can produce much better results with sharp edges. Nevertheless, it lacks the ability to recover clean HR image. In contrast, with a strong denoiser prior, DPIR produces the best visual result with both sharpness and naturalness.

\subsubsection{Intermediate Results and Convergence}
Fig.~\ref{fig:sr_iter}(a)-(e) provides the visual results and PSNR results of $\mathbf{x}_k$ and $\mathbf{z}_k$ at different iterations of DPIR on the testing image from Fig.~\ref{fig:sr}. One can observe that, although the LR image contains no noise, the the closed-form solution $\mathbf{x}_1$ would introduce severe structured noise. However, it has a better PSNR than that of RCAN.
After passing $\mathbf{x}_1$ through the DRUNet denoiser, such structured noise is removed as can be seen from $\mathbf{z}_1$. Meanwhile, the tiny textures and structures are smoothed out and the edges become blurry. Nevertheless, the PSNR is significantly improved and is comparable to that of MZSR.
As the number of iterations increases, $\mathbf{x}_6$ contains less structured noise than $\mathbf{x}_1$, while $\mathbf{z}_6$ recovers more details and sharper edges than $\mathbf{z}_1$. The corresponding PSNR convergence curves are plotted in Fig.~\ref{fig:sr_iter}(f), from which we can see that $\mathbf{x}_k$ and $\mathbf{z}_k$ converge quickly to the fixed point.

\begin{figure*}[!htbp]
\begin{center}\hspace{-0.25cm}
\subfigure[Ground-truth]{
\begin{overpic}[width=0.157\textwidth]{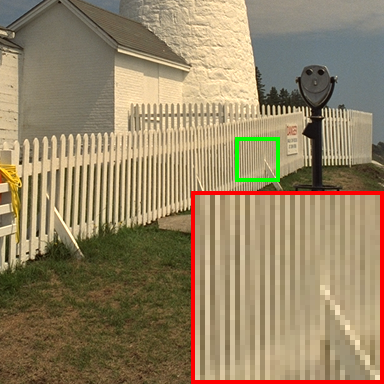}
\end{overpic}}
\subfigure[Matlab (33.67dB)]{
\begin{overpic}[width=0.157\textwidth]{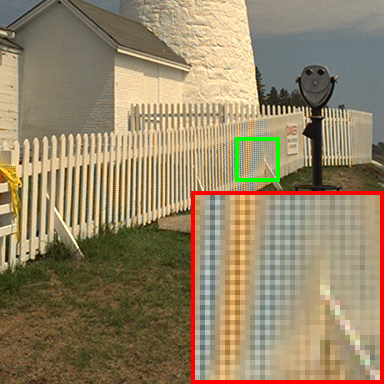}
\end{overpic}}
\subfigure[DDR (41.94dB)]{
\begin{overpic}[width=0.157\textwidth]{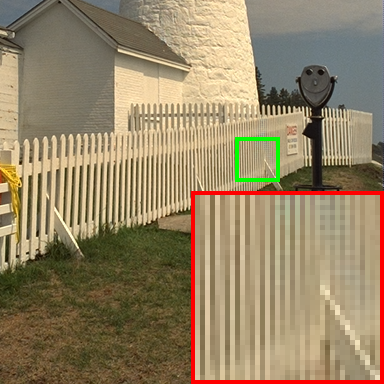}
\end{overpic}}
\subfigure[DeepJoint (42.49dB)]{
\begin{overpic}[width=0.157\textwidth]{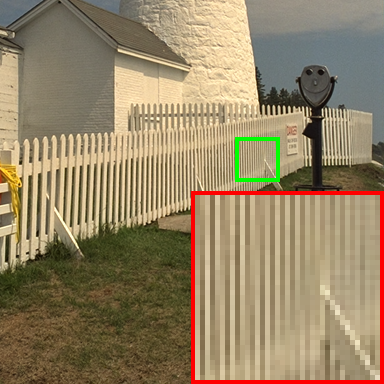}
\end{overpic}}
\subfigure[MMNet (40.62dB)]{
\begin{overpic}[width=0.157\textwidth]{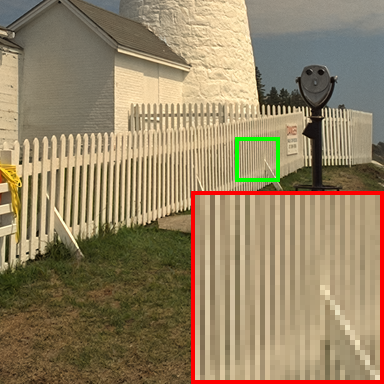}
\end{overpic}}
\subfigure[RNAN (43.77dB)]{
\begin{overpic}[width=0.157\textwidth]{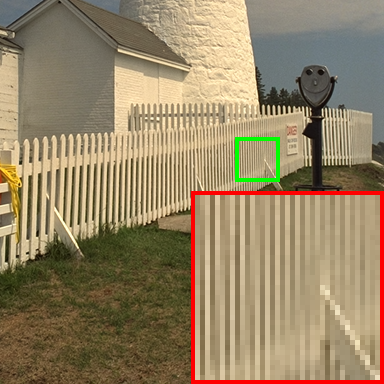}
\end{overpic}}
\subfigure[LSSC (42.31dB)]{
\begin{overpic}[width=0.157\textwidth]{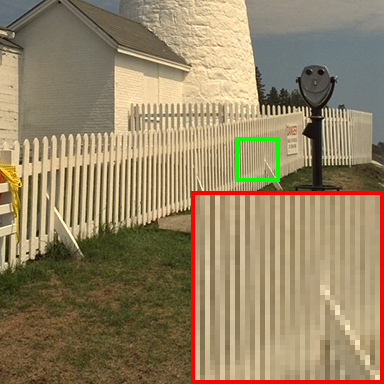}
\end{overpic}}
\subfigure[IRI (39.49dB)]{
\begin{overpic}[width=0.157\textwidth]{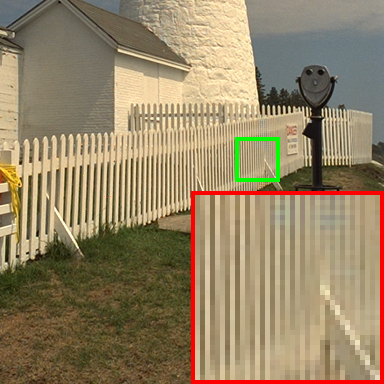}
\end{overpic}}
\subfigure[FlexISP (36.95dB)]{
\begin{overpic}[width=0.157\textwidth]{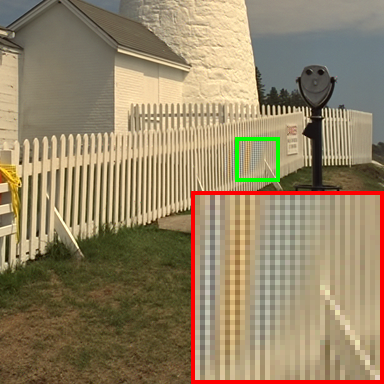}
\end{overpic}}
\subfigure[IRCNN (40.18dB)]{
\begin{overpic}[width=0.157\textwidth]{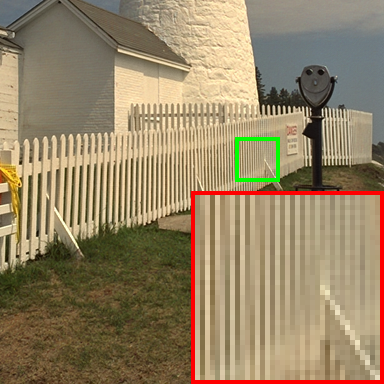}
\end{overpic}}
\subfigure[IRCNN+ (40.85dB)]{
\begin{overpic}[width=0.157\textwidth]{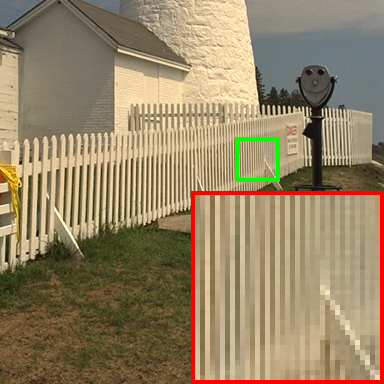}
\end{overpic}}
\subfigure[DPIR (43.23dB)]{
\begin{overpic}[width=0.157\textwidth]{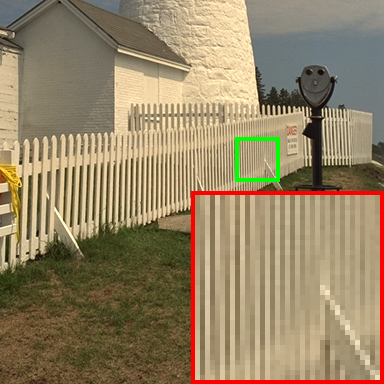}
\end{overpic}}
\caption{Visual results comparison of different demosaicing methods on image \textit{kodim19} from Kodak dataset.}\label{fig:dm}
\end{center}
\end{figure*}

\begin{figure*}[!htbp]
\centering\hspace{-0.2cm}
\subfigure[$\mathbf{x}_1$ (33.67dB)]{
\begin{overpic}[width=0.157\textwidth]{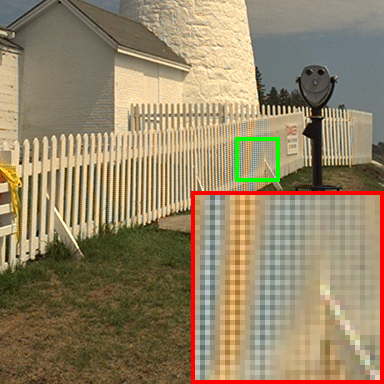}
\end{overpic}}
\subfigure[$\mathbf{z}_1$ (29.21dB)]{
\begin{overpic}[width=0.157\textwidth]{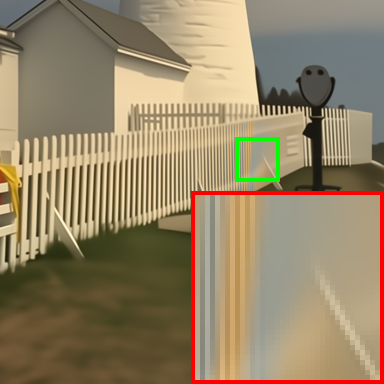}
\end{overpic}}
\subfigure[$\mathbf{x}_{16}$ (32.69dB)]{
\begin{overpic}[width=0.157\textwidth]{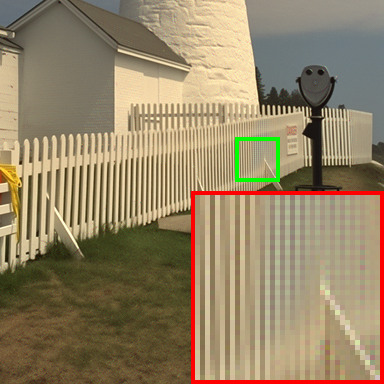}
\end{overpic}}
\subfigure[$\mathbf{z}_{16}$ (31.45dB)]{
\begin{overpic}[width=0.157\textwidth]{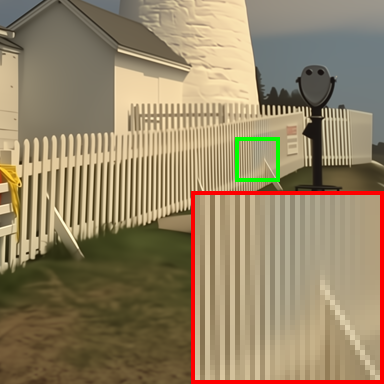}
\end{overpic}}
\subfigure[$\mathbf{x}_{40}$ (43.18dB)]{
\begin{overpic}[width=0.157\textwidth]{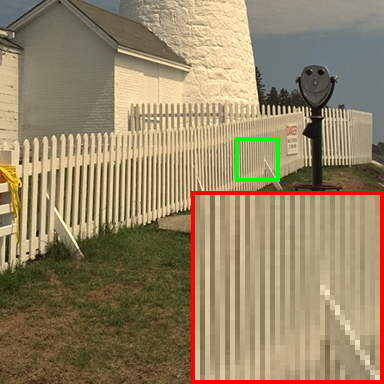}
\end{overpic}}
\subfigure[Convergence curves]{
\begin{overpic}[width=0.157\textwidth]{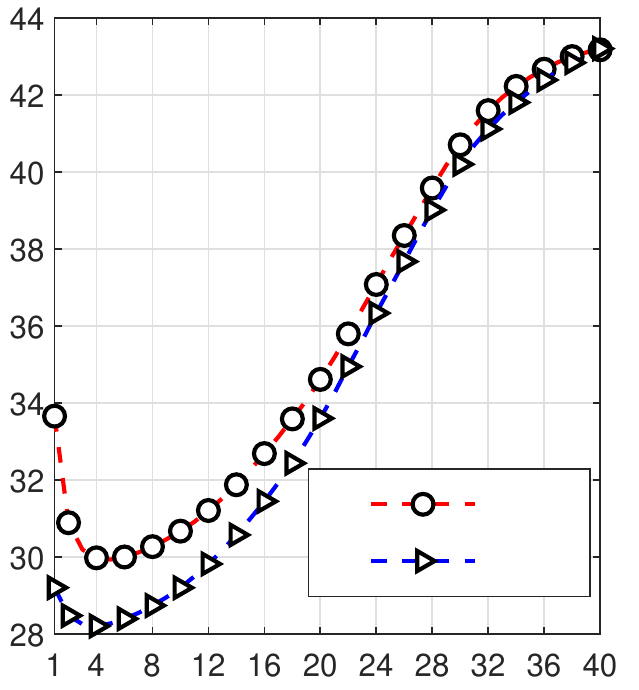}
\put(76,18){\color{black}{\tiny $\mathbf{z}_k$}}
\put(76,25.5){\color{black}{\tiny $\mathbf{x}_k$}}
\end{overpic}}
\vspace{-0.1cm}
\caption{(a)-(e) Visual results and PSNR results of $\mathbf{x}_k$ and $\mathbf{z}_k$ at different iterations; (f) Convergence curves of PSNR results (y-axis) for $\mathbf{x}_k$ and $\mathbf{z}_k$ with respect to number of iterations (x-axis).}%\vspace{-.1cm}
\label{fig:dm_iter}
\end{figure*}

\subsection{Color Image Demosaicing}

Current consumer digital cameras mostly use a single sensor with a color filter array (CFA) to record one of three R, G, and B values at each pixel location. As an essential process in camera pipeline, demosaicing aims to estimate the missing pixel values from a one-channel mosaiced image and the corresponding CFA pattern to recover a three-channel image~\cite{malvar2004high,gharbi2016deep,liu2020joint}.
The degradation model of mosaiced image can be expressed as
\begin{equation}\label{eq:deblur}
  \mathbf{y} = \mathbf{M}\odot\mathbf{x}
\end{equation}
where $\mathbf{M}$ is determined by CFA pattern and is a matrix with binary elements
indicating the missing pixels of $\mathbf{y}$, and $\odot$ denotes element-wise multiplication.
The closed-from solution of \eqref{eq33_1} is given by
\begin{equation}\label{eq_dm}
  \mathbf{x}_{k+1} = \frac{\mathbf{M}\odot\mathbf{y} + \alpha_k\mathbf{z}_{k}}{\mathbf{M} + \alpha_k}.
\end{equation}
In this paper, we consider the commonly-used Bayer CFA pattern with RGGB arrangement.
For the parameters $K$ and $\sigma_K$, they are set to 40 and 0.6, respectively. For $z_0$, it is initialized by matlab's \texttt{demosaic} function.

\subsubsection{Quantitative and Qualitative Comparison}

To evaluate the performance of DPIR for color image demosaicing, the widely-used Kodak dataset (consisting of 24 color images of size 768$\times$512) and McMaster dataset (consisting of 18 color images of size 500$\times$500) are used. The corresponding mosaiced images are obtained by filtering the color images with the Bayer CFA pattern.
 {The compared methods include
matlab's \texttt{demosaic} function~\cite{malvar2004high},
directional difference regression (DDR)~\cite{wu2016demosaicing},
deep unfolding majorization-minimization network (MMNet)~\cite{kokkinos2019iterative},
residual learning-based joint demosaicing-denoising (RLDD)~\cite{guo2020residual},
deep joint demosaicing and denoising (DeepJoint)~\cite{gharbi2016deep},
very deep residual non-local attention network (RNAN)~\cite{zhang2019residual}
learned simultaneous sparse coding (LSSC)~\cite{mairal2009non},
iterative residual interpolation (IRI)~\cite{ye2015color},
minimized-Laplacian residual interpolation (MLRI)~\cite{kiku2016beyond},
primal-dual algorithm with CBM3D denoiser prior (FlexISP)~\cite{heide2014flexisp},
IRCNN and IRCNN+. Note that DDR, MMNet, RLDD, DeepJoint, and RNAN are learning-based methods, while LSSC, IRI, MLRI, FlexISP, IRCNN, IRCNN+, and DPIR are model-based methods.}

Table~\ref{table_demosaicing} reports the average PSNR(dB) results of different methods on Kodak dataset and McMaster dataset.  It can be seen that while RNAN and MMNet achieve the best results, DPIR can have a very similar result and significantly outperforms the other model-based methods. With a stronger denoiser, DPIR has an average PSNR improvement up to 1.8dB over IRCNN+.

Fig.~\ref{fig:dm} shows the visual results comparison of different methods on a testing image from Kodak dataset. As one can see, the Matlab's simple demosaicing method introduces some zipper effects and false color artifacts. Such artifacts are highly reduced by learning-based methods such as DeepJoint, MMNet and RNAN. For the model-based methods, DPIR produces the best visual results whereas the others give rise to noticeable artifacts.

\subsubsection{Intermediate Results and Convergence}

Figs.~\ref{fig:dm_iter}(a)-(e) show the visual results and PSNR results of $\mathbf{x}_k$ and $\mathbf{z}_k$ at different iterations. One can see that the DRUNet denoiser prior plays the role of smoothing out current estimation $\mathbf{x}$. By passing $\mathbf{z}$ through \eqref{eq_dm}, the new output $\mathbf{x}$ obtained by a weighted average of $\mathbf{y}$ and $\mathbf{z}$ becomes unsmooth. In this sense, the denoiser also aims to diffuse $\mathbf{y}$ for a better estimation of missing values.
Fig.~\ref{fig:dm_iter}(f) shows the PSNR convergence curves of $\mathbf{x}_k$ and $\mathbf{z}_k$. One can see that the two PSNR sequences are not monotonic but they eventually converge to the fixed point. Specifically, a decrease of the PSNR value for the first four iterations can be observed as the denoiser with a large noise level removes much more useful information than the unwanted artifacts.

\section{Discussion}\label{discussion}

\begin{figure}[!htbp]
\centering
\subfigure{\includegraphics[width=0.158\textwidth]{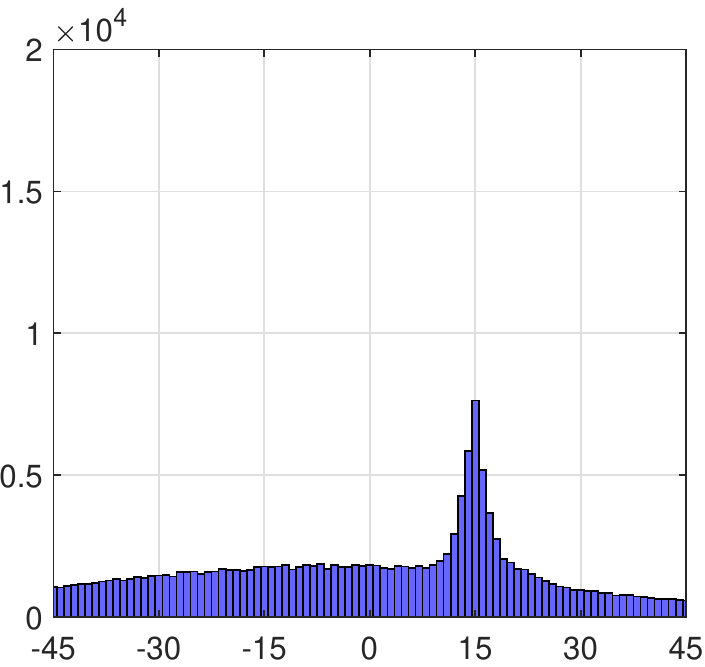}}
\subfigure{\includegraphics[width=0.158\textwidth]{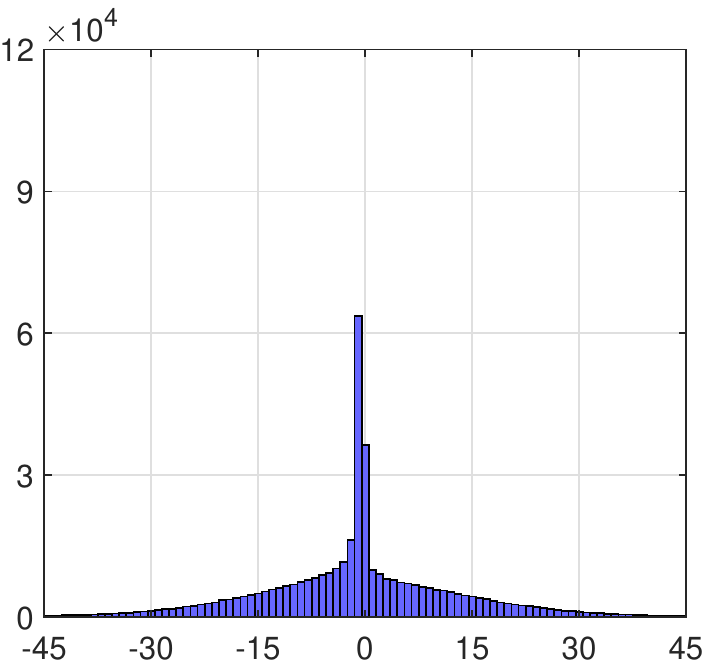}}
\subfigure{\includegraphics[width=0.158\textwidth]{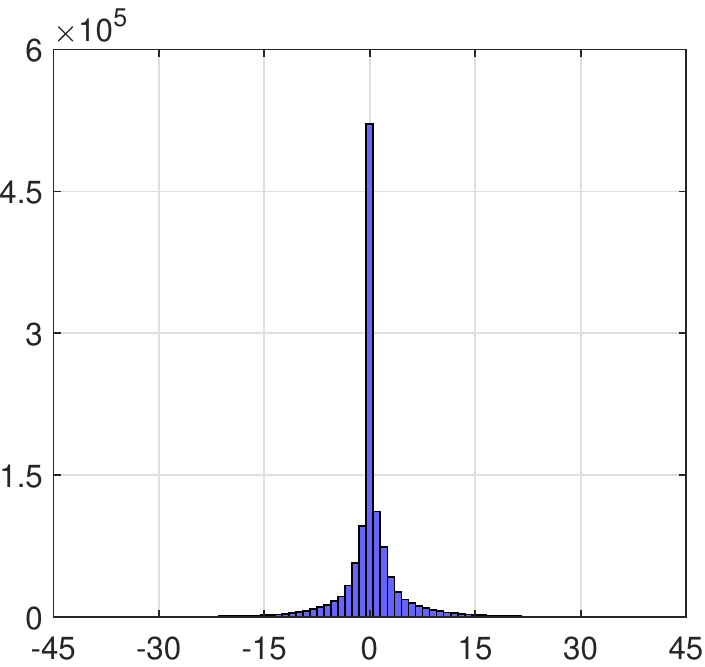}}\setcounter{subfigure}{0}
\subfigure[Deblurring]{\includegraphics[width=0.158\textwidth]{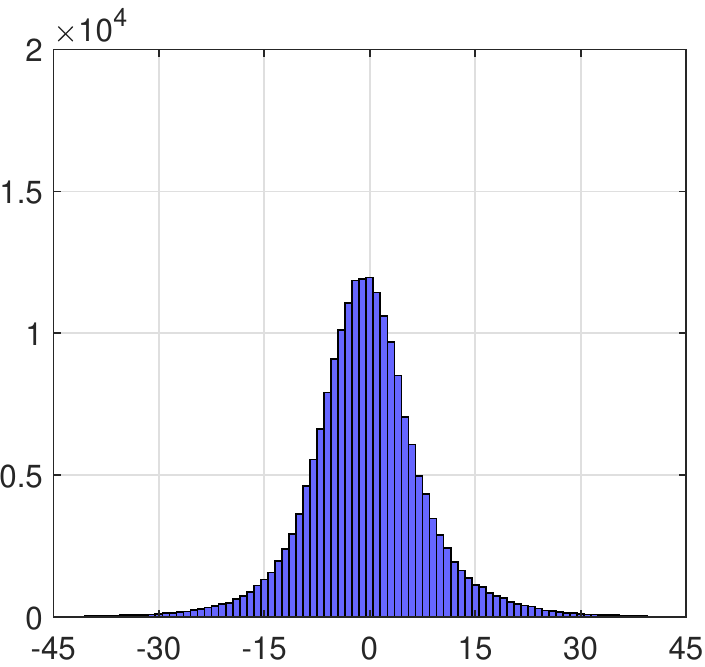}}
\subfigure[Super-Resolution]{\includegraphics[width=0.158\textwidth]{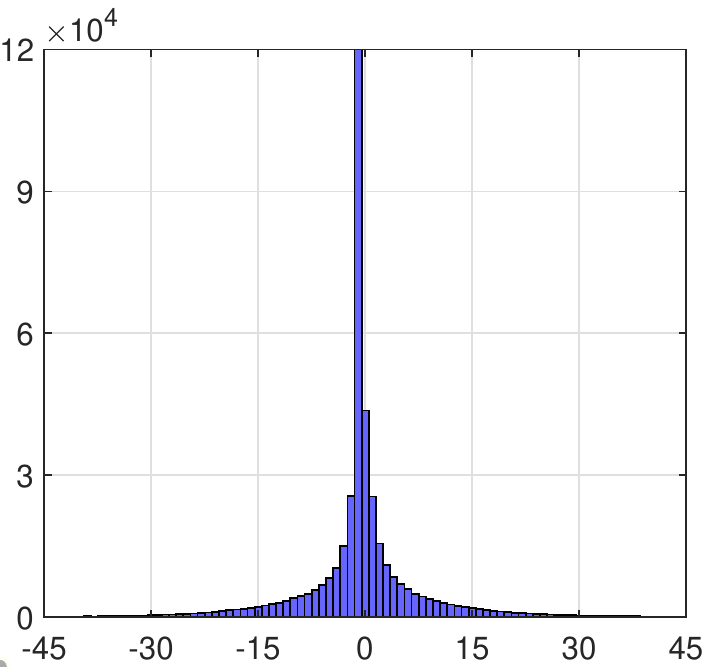}}
\subfigure[Demosaicing]{\includegraphics[width=0.158\textwidth]{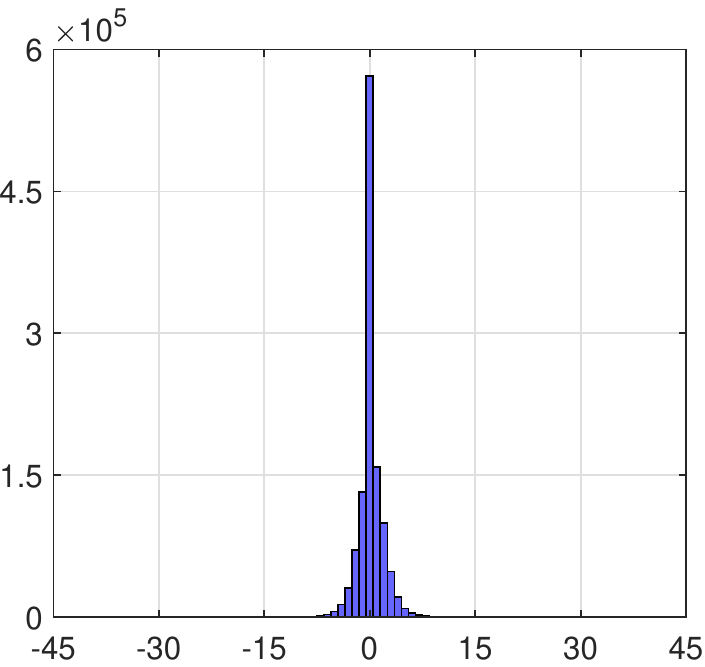}}
% \vspace{-0.1cm}
\caption{Histogram of the noise (difference) between the ground-truth and input of the denoiser in the first iteration (first row) and last iteration (second row) for (a) deblurring, (b) super-resolution, and (c) demosaicing. The histograms are based on $\mathbf{x}_1$ and $\mathbf{x}_8$ in Fig.~\ref{fig:deblur_iter}, $\mathbf{x}_1$ and $\mathbf{x}_{24}$ in Fig.~\ref{fig:sr_iter} and $\mathbf{x}_1$ and $\mathbf{x}_{40}$ in Fig.~\ref{fig:dm_iter}.}
\label{fig:iter}
\end{figure}

While the denoiser prior for plug-and-play IR is trained for Gaussian denoising, it does not necessary mean the noise of its input (or more precisely, the difference to the ground-truth) has a Gaussian distribution. In fact, the noise distribution varies across different IR tasks and even different iterations.
Fig.~\ref{fig:iter} shows the noise histogram of $\mathbf{x}_1$ and $\mathbf{x}_8$ in Fig.~\ref{fig:deblur_iter} for deblurring, $\mathbf{x}_1$ and $\mathbf{x}_{24}$ in Fig.~\ref{fig:sr_iter} for super-resolution, and $\mathbf{x}_1$ and $\mathbf{x}_{40}$ in Fig.~\ref{fig:dm_iter} for demosaicing.
It can be observed that the three IR tasks has very different noise distributions. This is intuitively reasonable because the noise also correlates with the degradation operation which is different for the three IR tasks.
Another interesting observation is that the two noise distributions of $\mathbf{x}_1$ and $\mathbf{x}_8$ in Fig.~\ref{fig:iter}(a) are different and the latter tends to be Gaussian-like. The underlying reason is that the blurriness caused by blur kernel is gradually alleviated after several iterations. In other words, $\mathbf{x}_8$ suffers much less from the blurriness and thus is dominated by Gaussian-like noise.

According to the experiments and analysis, it can be concluded that the denoiser prior mostly removes the noise along with some fine details, while the subsequent data subproblem plays the role of alleviating the noise-irrelevant degradation and adding the lost details back. Such mechanisms actually enable the plug-and-play IR to be a generic method. However, it is worth noting that this comes at the cost of losing efficiency and specialization because of such general-purpose Gaussian denoiser prior and the manual selection of hyper-parameters. In comparison, deep unfolding IR can train a compact inference with better performance by jointly learning task-specific denoiser prior and hyper-parameters.
Taking SISR as an example, rather than smoothing out the fine details by deep plug-and-play denoiser, the deep unfolding denoiser can recover the high-frequency details.

\section{Conclusion}

In this paper, we have trained flexible and effective deep denoisers for plug-and-play image restoration. Specifically, by taking advantage of half-quadratic splitting algorithm, the iterative optimization of three different image restoration tasks, including deblurring, super-resolution and color image demosaicing, consists of alternately solving a data subproblem which has a closed-form solution and a prior subproblem which can be replaced by a deep denoiser.
Extensive experiments and analysis on parameter setting, intermediate results, empirical convergence were provided. The results have demonstrated that plug-and-play image restoration with powerful deep denoiser prior have several advantages. On the one hand, it boosts the effectiveness of model-based methods due to the implicit but powerful prior modeling of deep denoiser. On the other hand, without task-specific training, it is more flexible than learning-based methods while having comparable performance.
In summary, this work has highlighted the advantages of deep denoiser based plug-and-play image restoration.
It is worth noting that there also remains room for further study. For example, one direction would be
how to integrate other types of deep image prior, such as deep generative prior~\cite{brock2018large}, for effective image restoration.

\section{Acknowledgements}
This work was partly supported in part by the ETH Z\"urich Fund (OK), in part by Huawei, in part by the National Natural Science Foundation of China under grant No. U19A2073, in part by Amazon AWS and in part by Nvidia grants.

\bibliographystyle{IEEEtran}
\bibliography{IEEEabrv,egbib}

\end{document}